\documentclass[12pt,a4paper]{article}

\usepackage{amsmath,amssymb}
\usepackage{psfrag}
\usepackage{graphicx}
\usepackage{slashed}

\makeatletter \@addtoreset{equation}{section} \makeatother

\makeatletter
\let\old@startsection=\@startsection
\let\oldl@section=\l@section
\renewcommand{\@startsection}[6]{\old@startsection{#1}{#2}{#3}{#4}{#5}{#6\mathversion{bold}}}
\renewcommand{\l@section}[2]{\oldl@section{\mathversion{bold}#1}{#2}}
\makeatother

\makeatletter
\let\old@makecaption=\@makecaption
\def\@makecaption{\small\old@makecaption}
\makeatother

\let\oldPhi=\Phi
\let\oldPsi=\Psi
\let\oldGamma=\Gamma
\let\oldDelta=\Delta
\let\oldSigma=\Sigma
\let\oldTheta=\Theta
\let\oldPi=\Pi
\renewcommand{\Phi}{\mathnormal{\oldPhi}}
\renewcommand{\Psi}{\mathnormal{\oldPsi}}
\renewcommand{\Gamma}{\mathnormal{\oldGamma}}
\renewcommand{\Sigma}{\mathnormal{\oldSigma}}
\renewcommand{\Delta}{\mathnormal{\oldDelta}}
\renewcommand{\Theta}{\mathnormal{\oldTheta}}
\renewcommand{\Pi}{\mathnormal{\oldPi}}

\newcommand{\Action}{\mathcal{S}}
\newcommand{\Lagr}{\mathcal{L}}

\newcommand{\tr}{\mathop{\mathrm{tr}}}

\newcommand{\sign}{\mathop{\mathrm{sign}}}

\renewcommand{\Im}{\mathop{\mathrm{Im}}}

\newcommand{\order}[1]{\mathcal{O}(#1)}

\newcommand{\Reals}{R}   
\newcommand{\Sphere}{S}  
\newcommand{\AdS}{\mathrm{AdS}}

\ifx\genfrac\sdflkaj

\else

\fi
\newcommand{\sfrac}[2]{{\textstyle\frac{#1}{#2}}}
\newcommand{\half}{\sfrac{1}{2}}
\newcommand{\ihalf}{\sfrac{i}{2}}
\newcommand{\quarter}{\sfrac{1}{4}}

\newcommand{\matr}[2]{\left(\begin{array}{#1}#2\end{array}\right)}
\newcommand{\alg}[1]{\mathfrak{#1}}
\newcommand{\grp}[1]{\mathrm{#1}}

\newcommand{\grSO}{\grp{SO}}
\newcommand{\algSU}{\alg{su}}

\newcommand{\brk}[1]{(#1)}
\newcommand{\lrbrk}[1]{\left(#1\right)}
\newcommand{\bigbrk}[1]{\bigl(#1\bigr)}
\newcommand{\Bigbrk}[1]{\Bigl(#1\Bigr)}

\newcommand{\lrsbrk}[1]{\left[#1\right]}
\newcommand{\bigsbrk}[1]{\bigl[#1\bigr]}
\newcommand{\Bigsbrk}[1]{\Bigl[#1\Bigr]}

\newcommand{\Biggsbrk}[1]{\Biggl[#1\Biggr]}
\newcommand{\ket}[1]{\mathopen{|}#1\mathclose{\rangle}}
\newcommand{\bra}[1]{\mathopen{\langle}#1\mathclose{|}}
\newcommand{\braket}[2]{\mathopen{\langle}#1|#2\mathclose{\rangle}}

\newcommand{\comm}[2]{[#1,#2]}

\newcommand{\acomm}[2]{\{#1,#2\}}

\newcommand{\abs}[1]{{|#1|}}

\newcommand{\pint}{\makebox[0pt][l]{\hspace{3.4pt}$-$}\int}

\def\[{\begin{equation}}
\def\]{\end{equation}}
\def\<{\begin{eqnarray}}
\def\>{\end{eqnarray}}

\makeatletter
\def\mr@ignsp#1 {\ifx\:#1\@empty\else #1\expandafter\mr@ignsp\fi}%
\newcommand{\multiref}[1]{\begingroup
\xdef\mr@no@sparg{\expandafter\mr@ignsp#1 \: }%
\def\mr@comma{}%
\@for\mr@refs:=\mr@no@sparg\do{\mr@comma\def\mr@comma{,}\ref{\mr@refs}}%
\endgroup}
\makeatother

\newcommand{\hypref}[2]{\ifx\href\asklfhas #2\else\href{#1}{#2}\fi}

\newcommand{\secref}[1]{Sec.~\multiref{#1}}

\newcommand{\appref}[1]{App.~\multiref{#1}}

\newcommand{\figref}[1]{Fig.~\multiref{#1}}
\renewcommand{\eqref}[1]{(\multiref{#1})}

\newcommand{\bibtitle}[1]{\emph{#1}}
\newcommand{\hepth}[1]{\texttt{hep-th/#1}}
\newcommand{\condmat}[1]{\texttt{cond-mat/#1}}

\ifx\href\asklfhas\newcommand{\href}[2]{#2}\fi
\newcommand{\arxivno}[1]{\href{http://arxiv.org/abs/#1}{#1}}

\newcommand{\pfi}{\varphi}

\newcommand{\bpsi}{\bar{\psi}}

\newcommand{\spa}[1]{\partial_x{#1}}

\newcommand{\comma}{\quad,\quad}
\newcommand{\unit}{\mathbf{1}}

\newcommand{\vx}{\vec{x}}
\newcommand{\vp}{\vec{p}}
\newcommand{\vpp}{\vec{p}\,'}
\newcommand{\vk}{\vec{k}}
\newcommand{\vkp}{\vec{k}'}
\newcommand{\vq}{\vec{q}}

\newcommand{\Proj}{\mathbb{P}}

\newcommand{\momintt}[1]{\int \! \frac{d^2 #1}{(2\pi)^2}\:}

\begin{document}
\setcounter{page}{0}

\thispagestyle{empty}
\begin{flushright}\footnotesize
\texttt{\arxivno{hep-th/0603039}}\\
\texttt{ITEP-TH-07/06}\\
\texttt{UUITP-02/06}\\
\vspace{0.5cm}
\end{flushright}
\vspace{0.5cm}

\renewcommand{\thefootnote}{\fnsymbol{footnote}}
\setcounter{footnote}{0}

\begin{center}
{\Large\textbf{\mathversion{bold}
Bethe Ansatz in Stringy Sigma Models}\par} \vspace{1cm}

\textsc{T.~Klose and K.~Zarembo\footnote{Also at ITEP, Moscow,
117259 Bol. Cheremushkinskaya 25, Moscow, Russia}} \vspace{5mm}

\textit{Department of Theoretical Physics, Uppsala University\\
P.O.~Box 803, SE-75108, Uppsala, Sweden}\vspace{3mm}

\texttt{thomas.klose,konstantin.zarembo@teorfys.uu.se}\\
\par\vspace{1cm}

\vfill

\textbf{Abstract}\vspace{5mm}

\begin{minipage}{12.7cm}
We compute the exact S-matrix and give the Bethe ansatz solution for
three sigma-models which arise as subsectors of string theory in
$\AdS_5\times \Sphere^5$: Landau-Lifshitz model (non-relativistic
sigma-model on $\Sphere^2$), Alday-Arutyunov-Frolov model (fermionic
sigma-model with $\algSU(1|1)$ symmetry), and Faddeev-Reshetikhin
model (string sigma-model on $\Sphere^3\times \Reals$).
\end{minipage}

\vspace*{\fill}

\end{center}

\newpage
\setcounter{page}{1}
\renewcommand{\thefootnote}{\arabic{footnote}}
\setcounter{footnote}{0}

\tableofcontents
\vspace{10mm}
\hrule
\vspace{10mm}

\section{Introduction}
\label{sec:intro}

According to the AdS/CFT correspondence \cite{Maldacena:1998re}
solving four-dimensional $\mathcal{N}=4$ super-Yang-Mills (SYM)
theory amounts to quantization of type IIB superstrings on the
$\AdS_5\times \Sphere^5$ background. This potentially easier problem
has so far resisted solution, partly because the sigma-model on
$\AdS_5\times \Sphere^5$ \cite{Metsaev:1998it} is of the
Green-Schwarz type and has well-known difficulties in the conformal
gauge. The sigma model, however, is integrable \cite{Bena:2003wd}
which gives us hope to solve it with the help of Bethe ansatz. There
is a mounting evidence that the non-perturbative spectrum of AdS/CFT
is indeed described by some sort of Bethe equations
\cite{Beisert:2004ry,Zarembo:2004hp}. The strongest evidence comes
from studying the spectrum of anomalous dimensions in the SYM theory
\cite{Minahan:2002ve,Beisert:2003tq,Beisert:2005fw} and from
analyzing classical solutions in the sigma-model
\cite{Kazakov:2004qf,Beisert:2005bm,Dorey:2006zj}, which describe
spinning strings in $\AdS_5\times \Sphere^5$ \cite{Gubser:2002tv}.

The Bethe ansatz \cite{Bethe:1931hc,Faddeev:1996iy,Korepin_book} is
a common method to solve integrable models, which encodes the whole
spectrum of the system in a set of algebraic equations. The Bethe
equations can be interpreted in terms of factorized particle
scattering, where the ``Bethe particles'' do not necessarily
coincide with the physical degrees of freedom of the theory. In such
an interpretation, the Bethe equations arise as quantization
conditions for the momenta of the particles in a box of linear size
$L$:
\[ \label{BEmaster}
 e^{ip_jL} = \prod_{k\neq j}^{} \, e^{-i\Delta(p_j,p_k)} \; ,
\]
where $\Delta (p_j,p_k)$ is the phase shift experienced by the $j$th
particle when it scatters off the $k$th particle. The energy of a
Bethe state is the sum of the single-particle energies:
\[
 E_{\{p_j\}}=\sum_{j}^{}\varepsilon (p_j) \; .
\]
The energy spectrum is thus completely determined by the  two-body
scattering and the one-body dispersion relation. There are no
genuine multi-body interactions \cite{zz:factorized-s-matrix}.

The SYM theory is an example where the Bethe particles are very
different from the excitations in the four-dimensional space-time.
Here, the Bethe equations, which determine the planar spectrum of
anomalous dimension, were derived by interpreting single trace
operators in SYM as quantum states of abstract spin chains
\cite{Minahan:2002ve}. The Bethe particles are waves that propagate
along the spin-chain associated with a  local operator in the SYM,
but more recent results \cite{Rej:2005qt} suggest that the story
might be different (and far more interesting!) at the
non-perturbative level. On the string-theory side, the evidence in
favor of the Bethe-ansatz structure of the spectrum comes from the
two sources: (i) The classical solutions of the string sigma-model
can be parameterized by the integral equations of Bethe type
\cite{Kazakov:2004qf,Beisert:2005bm} and (ii) the leading quantum
corrections in the near-BMN limit of the $\AdS_5\times \Sphere^5$
geometry \cite{Callan:2003xr} can be parameterized by a set Bethe
equations \cite{Arutyunov:2004vx,Staudacher:2004tk}. The quantum
string Bethe equations, which were conjectured on the basis of these
two observations
\cite{Arutyunov:2004vx,Staudacher:2004tk,Beisert:2005fw,Frolov:2006cc},
receive corrections at higher orders in the sigma-model coupling
\cite{Beisert:2005cw} and thus contain infinitely many unknown
parameters. Deriving  Bethe equations for quantum string in
$\AdS_5\times \Sphere^5$ from first principles is an open problem.
In particular it is not quite clear what degrees of freedom of the
string are represented by the Bethe particles.

The string sigma-model on $\AdS_5\times \Sphere^5$ is a rather
complicated two-dimensional field theory. Similar but simpler
systems, such as the $\grp{Osp}(2m+2|2m)$ coset sigma-model, were
solved by Bethe ansatz \cite{Mann:2005ab}, because many
two-dimensional factorized S-matrices are known exactly
\cite{zz:factorized-s-matrix}. For other models a direct
relationship to spin chains can be established
\cite{Polyakov:2005ss}. The idea that we would like to put forward
in this paper, and test on a number of simplified models, is to
derive the Bethe equations for quantum strings by explicitly
computing the two-body scattering matrix on the world-sheet. We use
relatively simple and rather standard methods of quantum field
theory to do that. This will allow us to obtain the Bethe equations
for several two-dimensional field theories that arise as reductions
of the $\AdS_5\times \Sphere^5$ sigma-model. We should mention that
the S-matrix approach was quite successfully used on the
gauge-theory side of the AdS/CFT correspondence
\cite{Staudacher:2004tk,Freyhult:2005ws,Beisert:2005tm}.

We demand (rather than prove) quantum integrability and
factorization of the S-matrix as necessary prerequisites. With these
assumptions in mind, our method provides the means to \emph{derive}
the Bethe equations of an integrable theory at the quantum level. A
more rigorous approach to Bethe ansatz is the quantum inverse
scattering method \cite{Faddeev:1979gh} which fully exploits the
rich algebraic structure associated with integrability.

The first model we consider is a non-relativistic sigma model  on
$\Sphere^2$ (the Landau-Lifshitz model). This model describes fast
moving strings on $\Sphere^3\times \Reals$ and arises as a
low-energy effective theory of the Heisenberg ferromagnet (the
connection that plays an important role in the AdS/CFT
correspondence \cite{Kruczenski:2003gt}).
We derive the Bethe equations for the LL model in \secref{sec:ll} by
direct computation of the S-matrix. In \secref{sec:aaf} we consider
the fermionic sigma-model that arises from the $\algSU(1|1)$
reduction of the AdS string \cite{aaf:su11-string}. Finally, in
\secref{sec:fr} we consider again string theory on $\Sphere^3\times
\Reals$, but this time without making the low energy approximation.
This model was introduced by Faddeev and Reshetikhin
\cite{fr:principle-chiral-field} and is closely related to the
$\algSU(2)$ principal chiral field, whose Bethe-ansatz solution was
obtained in \cite{Polyakov:1983tt}. The lattice-regularized quantum
version of the FR model was solved in
\cite{fr:principle-chiral-field}. We derive the Bethe equations
directly in the continuum.

\section{Preliminaries}
\label{sec:prelim}

The general idea is to derive the Bethe equations for quantum strings from scattering computations in the world-sheet theory.
Let us sketch this idea for the example of a single scalar field on the world-sheet. In the usual Hamiltonian approach to the Bethe equations, one constructs eigenstates which have the structure of scattering plane-waves. In the two-particle sector such an eigenstate of the Hamiltonian looks like
\[ \label{eqn:two-particle-eigenstate}
  \ket{p\, p'}
  = \int dx\, dx' \,
  \underbrace{
  \bigsbrk{ \theta(x'-x) + \theta(x-x') S(p,p') } \,
  e^{i p x + i p' x'} }_{\chi(x,x')} \,
  \pfi^\dag(x) \pfi^\dag(x') \,
  \ket{0} \; .
\]
It is parameterized by two momenta $p$ and $p'$. We will always
label the momenta such that $p > p'$, so that the first term in the
wave function is the incoming wave, and the second term is the
scattered wave. Imposing the periodicity condition on the wave
function: $\chi(0,x') = \chi(L,x')$ and $\chi(x,0) = \chi(x,L)$, one
finds that the momenta of the particles are quantized according to
\[
  e^{i L p} = S(p',p) \comma e^{i L p'} = S(p,p') \; ,
\]
where $S(p',p) = 1/S(p,p')$. This is a particular case of
\eqref{BEmaster}, for which integrability is
actually not required. The distinguishing feature of integrable
models is that the multi-body wave function is two-particle
reducible \cite{Thacker:1980ei,Korepin_book}. The periodicity
condition for an arbitrary multi-particle eigenstate of the
Hamiltonian is then expressed in terms of the two-body phase shifts
as in \eqref{BEmaster}.

Diagonalization of the Hamiltonian, however, is not the most
efficient way to calculate the scattering phase shifts in field
theory. It is much easier to compute $S(p,p')$ as the matrix element
of the infinite time evolution operator $\hat{S}$ between the
two-particle scattering states:
\[ \label{eqn:S-matrix}
  \bra{k\, k'} \hat{S} \ket{p\, p'} = S(p,p') \, \delta_+(p,p',k,k') \; .
\]
Here we have introduced the notation
\[ \label{eqn:emc}
  \delta_\pm(p,p',k,k') = (2\pi)^2 \bigbrk{ \delta(p-k) \delta(p'-k') \pm \delta(p-k') \delta(p'-k) } \; .
\]
This factor represents  the conservation of individual momenta
during the scattering process. In two dimensions, energy and
momentum conservation allows two particles only to exchange their momenta.
The relative sign between the two terms is plus for bosons and minus
for fermions.

We will compute the S-matrix  using Feynman diagrams. Note that the
usual Feynman rules calculate the matrix element
$\mathcal{M}(p,p',k,k')$ as defined in
\[
  \bra{k\, k'} (\hat{S}-\unit) \ket{p\, p'} = i\mathcal{M}(p,p',k,k')
  \, (2\pi)^2 \, \delta^{(2)}(p^\mu+{p'}^\mu-k^\mu-{k'}^\mu) \; .
\]
The energy-momentum conserving delta-function is different from
\eqref{eqn:emc}  by a Jacobian $1/(\partial \varepsilon/\partial p -
\partial\varepsilon/\partial p')$, which has to be taken into
account when extracting the phase shift from the diagrammatic
calculations.

In what follows we make use of  \eqref{eqn:S-matrix} and derive the
Bethe equations for three different theories that are of potential
relevance for quantum strings in  $\AdS_5\times \Sphere^5$.

\section{Landau-Lifshitz model}
\label{sec:ll}

The LL model is defined by the action
\begin{equation}\label{LLaction}
\Action = \int d^2x \left[ C_t(\vec{n}) -
\frac{1}{4}(\spa{\vec{n}})^2 \right] \; ,
\end{equation}
where $\vec{n}$ is a three-dimensional unit vector:
\begin{equation}\label{constra}
 \vec{n}^2=1 \; .
\end{equation}
 The action contains the non-local Wess-Zumino term
\begin{equation}\label{eqn:non-local-WZ}
 C_q(\vec{n})=-\frac{1}{2}\int_{0}^{1}d\xi \,\varepsilon _{ijk}\,n_i\,
 \partial _\xi n_j\,\partial _q n_k \; ,
\end{equation}
and is of the first order in time derivatives. The equations of
motion that follow from  (\ref{LLaction}) are
\begin{equation}\label{LLeqm}
 \partial _tn_i=\varepsilon _{ijk}\,n_j\,\partial^2 _xn_k \; .
\end{equation}
The LL equation is completely integrable \cite{LLclass}. The quantum
inverse scattering method for the LL model was discussed in
\cite{Sklyanin}.

The LL model is the low-energy effective field theory of the
Heisenberg ferromagnet with Hamiltonian
\begin{equation}\label{Werner}
 \Gamma =\frac{\lambda }{16\pi ^2}\sum_{l=1}^{L}\left(1-\vec{\sigma }_l
 \cdot\vec{\sigma }_{l+1}\right) \; ,
\end{equation}
which in the AdS/CFT context arises as the one-loop mixing matrix of
scalar composite operators $\tr (Z^{L-M}W^M+{{\rm permutations}})$
\cite{Minahan:2002ve}. The Heisenberg equations of motion for
(\ref{Werner}) reduce to (\ref{LLeqm}) in the continuum limit
$L\rightarrow \infty $ if the spin operators are formally replaced
by unit c-number vectors: $\vec{\sigma }_l(\tau )\rightarrow
\vec{n}(\tau ,\sigma )$ with $\sigma =2\pi l/L$. Alternatively, the
action in the coherent-state path integral of the Heisenberg model
in the continuum limit becomes \cite{Kruczenski:2003gt}
\begin{equation}\label{}
 \Action=\frac{L}{2\pi }\int_{}^{}d\tau \int_{0}^{2\pi }d\sigma \,
 \left[C(\vec{n})-\frac{\lambda }{8L^2}
 \left(\partial _\sigma \vec{n}\right)^2\right].
\end{equation}
 This is the same as (\ref{LLaction}) after
the following rescalings: $\sigma =2\pi x/L$, $\tau =8\pi
^2t/\lambda $. The spacial coordinate $x$ in (\ref{LLaction}) now
has periodicity $L$.

The LL model can be also derived from classical string theory on
$\Sphere^3\times \Reals$ in the limit of fast-moving strings. The
details of the derivation, together with the precise matching  to
the mixing matrix for long operators in the SYM, can be found in
\cite{Kruczenski:2003gt}. Here we would like to view the LL model as
a (1+1)-dimensional quantum field theory.
We derive the Bethe equations for the non-perturbative spectrum of
the LL model with the help of relatively simple perturbative
calculations, which are similar to the perturbative calculation of
the S-matrix in the closely non-linear Schr\"odinger model
\cite{Thacker:1974kv}. Quantum-mechanical (Hamiltonian) perturbation
theory for the LL model was developed in
\cite{Minahan:2005mx,Minahan:2005qj}. Here, we will use Feynman diagrams
which greatly facilitates the calculation of the S-matrix.

In order to develop perturbation theory we first re-write the WZ
term in the local form. For that we will need some properties of the
WZ action, which we review here following \cite{Faddeev's_book}. The
WZ term in (\ref{LLaction}) can be written as
\begin{equation}\label{}
 {\rm WZ}[\vec{n}] := \int\!dt\: C_t(\vec{n}) = -\frac{1}{4}\int_{}^{}\varepsilon _{ijk}\,n_idn_j\wedge
 dn_k \; ,
\end{equation}
where $d\vec{n}=\partial _t\vec{n}\,dt+\partial _\theta
\vec{n}\,d\theta $. A short calculation shows that
\[
 \frac{1}{2}\,\varepsilon _{ijk}\,n_idn_j\wedge
 dn_k=\frac{dn_1\wedge
 dn_2}{n_3}=d\left(\frac{n_1dn_2-n_2dn_1}{1+n_3}\right) \; .
\]
This identity allows one to write the WZ action in a local form,
which comes at the price of losing manifest $\grSO(3)$ invariance:
\begin{equation}\label{WZlocal}
 {\rm WZ}[\vec{n}]=\frac{1}{2}\int_{}^{}dt\:\frac{\dot{n}_1n_2-\dot{n}_2n_1}{1+n_3} \; .
\end{equation}
The next step is to solve the constraint (\ref{constra}) by
expressing $n_3$ in term of $n_1$ and $n_2$. It is also convenient
to make a field  redefinition which gets rid of the non-linearities
in the kinetic term \cite{Minahan:2005mx}, and also to combine
$n_1$, $n_2$ into a single complex scalar, since the complex field
has the canonical non-relativistic propagator. The significance of
this fact will become clear later. So we define
\[\label{}
 \pfi =\frac{n_1+in_2}{\sqrt{2+2n_3}}\comma n_3=1-2\abs{\pfi}^2 \; .
\]
Performing this change of variables in (\ref{WZlocal}),
(\ref{LLaction}), we find:
\[\label{LLscalar}
\begin{split}
 \Action = \int_{}^{}d^2x\,\Biggsbrk{ &
 \frac{i}{2}\lrbrk{\pfi^* \partial_t \pfi - \partial_t \pfi^* \pfi}
   -\abs{\partial_x \pfi}^2
   -\frac{1}{4}\,\frac{2-\abs{\pfi}^2}{1-\abs{\pfi}^2}
    \lrsbrk{(\pfi^*\partial_x\pfi)^2+(\partial _x \pfi^* \pfi)^2} \\
 & -\frac{1}{2}\,\frac{\abs{\pfi}^4\abs{\partial _x\pfi}^2}{1-\abs{\pfi}^2}
   } \; .
\end{split}
\]
This action describes an interacting field theory of a single scalar
field. Its non-relativistic character leads to some important
non-renormalization properties.

First of all, the ground state is annihilated by the field operator:
\begin{equation}\label{}
 \pfi(t,x) \ket{0} = 0 \; .
\end{equation}
Since the equations of motion are of the first order in time
derivatives, the field operator in  the interaction picture
is expanded in negative-frequency modes only:
\[
  \pfi(t,x)  = \int \frac{dp}{2\pi} \, a_p  \,e^{-ip^2 t + ipx}
  \comma
  \pfi^*(t,x) = \int \frac{dp}{2\pi} \, a^\dag_p \,e^{ip^2 t - ipx} \; ,
\]
where $a_p$, $a^\dagger_p$ create and annihilate a particle with
momentum $p$ and energy\footnote{Recall that we have rescaled the
time variable by $8\pi ^2/\lambda$. Hence, the energy is the
frequency multiplied by $\lambda /8\pi ^2$}
\begin{equation}\label{disper}
  \varepsilon (p)=\frac{\lambda}{8\pi^2} p^2 \; .
\end{equation}
The operators $a_p$ and $a_p^\dagger $ obey canonical commutation
relation normalized as
\[
  \comm{a_p}{a^\dag_{p'}} = 2\pi\,\delta(p-p') \; .
\]

Since the ground state is annihilated by $\varphi(t,x)$, the
particles do not travel backwards in time and the propagator,
accordingly, has only one pole in the momentum representation:
\[ \label{eqn:LL-propagator}
\psfrag{xa}[c][c]{\small $(0,0)$}
\psfrag{xc}[c][c]{\small $(t,x)$}
\begin{split}
 D(t,x)
 & = \bra{0} T \,\pfi(t,x) \pfi^*(0,0) \ket{0}
   = \quad \raisebox{-6mm}{\includegraphics*{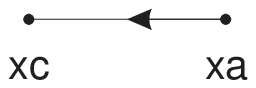}} \\
 & = \int\frac{d\omega\, dp}{(2\pi)^2} \,
 \frac{i}{\omega -  p^2 + i\epsilon} \, \,e^{-i \omega t +
 ipx} \; .
\end{split}
\]
The pole prescription, see \figref{fig:LLcontour}, or again the fact
that $\varphi(t,x)$ annihilates the vacuum makes the
coordinate-space propagator purely retarded:
\begin{figure}[t]
\begin{center}
\psfrag{neg}{$t<t'$}
\psfrag{pos}{$t>t'$}
\psfrag{ww}{$\omega$}
\psfrag{wp}[l][l]{$p^2 - i\epsilon$}
\includegraphics*{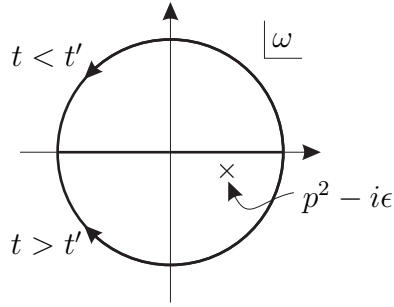}
\end{center}
\caption{\textbf{Pole prescription in Landau-Lifshitz model.}}
\label{fig:LLcontour}
\end{figure}
\[
 D(t,x)
= \theta(t)\sqrt{\frac{\pi }{it}}\,e^{\frac{ix^2}{4t}} \; .
\]

\begin{figure}[t]
\begin{center}
\psfrag{xa}[c][c]{$(x_1,t_1)$}
\psfrag{xb}[r][r]{$(x_2,t_2)$}
\psfrag{xc}[l][l]{$(x_3,t_3)$}
\psfrag{xd}[r][r]{\small $(x_4,t_4)$}
\psfrag{xn}[l][l]{\small $(x_n,t_n)$}
\psfrag{EPa}[l][l]{$(E_1,P_1)$}
\psfrag{EPb}[l][l]{$(E_2,P_2)$}
\psfrag{EPc}[l][l]{$(E_3,P_3)$}
\psfrag{EPd}[r][r]{\small $(E_4,P_4)$}
\psfrag{EPn}[r][r]{\small $(E_n,P_n)$}
\psfrag{Da}[l][l]{$D_{21}$}
\psfrag{Db}[l][l]{$D_{32}$}
\psfrag{Dc}[c][c]{$D_{42}$}
\psfrag{Dn}[r][r]{$D_{1n}$}
\psfrag{zero}{$=0$}
\includegraphics*{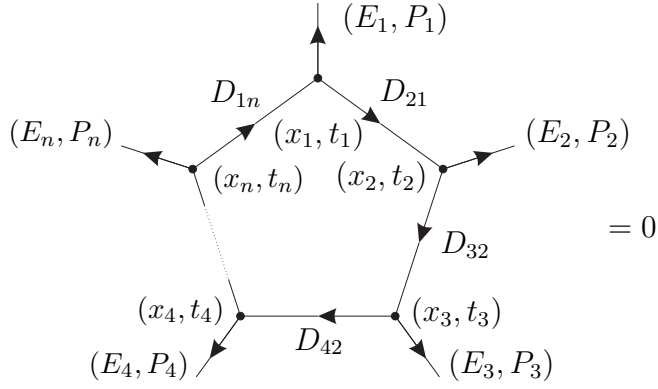}
\end{center}
\caption{\textbf{Non-renormalization theorem.} Any closed loop of likewise oriented propagators $D_{mn} \equiv D(t_m-t_n,x_m-x_n)$ vanishes.}
\label{fig:nonrenorm}
\end{figure}

\begin{figure}[t]
\begin{center}
\psfrag{p1}[c][c]{$p$} \psfrag{p2}[c][c]{$p'$}
\psfrag{k1}[c][c]{$k$} \psfrag{k2}[c][c]{$k'$}
\includegraphics*{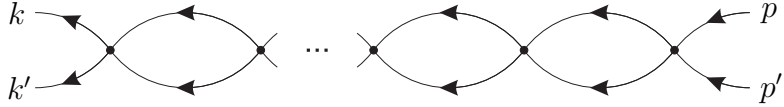}
\end{center}
\caption{\textbf{Generic loop diagram for the two-body S-matrix.}}
\label{fig:LL-loops}
\end{figure}

We can now prove the following ``non-renormalization theorem'': any
diagram that contains a closed loop with arrows in the same
direction vanishes, cf. \figref{fig:nonrenorm}.
In the coordinate representation this follows from the fact that at
least one propagator in the loop has a negative time argument.
Consequently,
\[
 D(t_2-t_1,x_2-x_1)D(t_3-t_2,x_3-x_2) \cdots D(t_1-t_n,x_1-x_n) = 0 \; .
\]
In the momentum space representation, the integrand has poles only in the lower half-plane of complex $\omega$.
\begin{figure}[t]
\begin{center}
\psfrag{ww}{$\omega$}
\includegraphics*{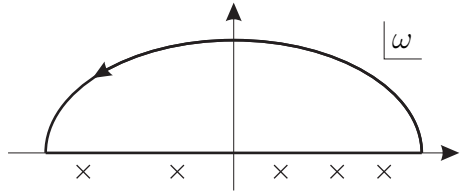}
\end{center}
\caption{\textbf{Poles for a closed loop.} All propagators in a closed have their poles in the lower half plane. Hence the integral \protect\eqref{eqn:energy-integral} over the energy flowing around the loop vanishes.}
\label{fig:nonrenorm-poles}
\end{figure}
The integral is then zero by the contour argument, cf. \figref{fig:nonrenorm-poles}:
\begin{equation}\label{eqn:energy-integral}
 \int_{}^{}\frac{d\omega }{2\pi }\:
 \frac{i}{\omega    -p^2      +i\epsilon} \,
 \frac{i}{\omega-E_1-(p-P_1)^2+i\epsilon} \,\ldots\,
 \frac{i}{\omega+E_n-(p+P_n)^2+i\epsilon}=0 \; .
\end{equation}
This theorem has three important consequences:
\begin{itemize}
\item The ground state energy is not renormalized: $E_{\rm vac}=0$.
This is consistent with the fact that the ferromagnetic vacuum is
the exact zero-energy eigenstate of the Heisenberg Hamiltonian
(\ref{Werner}) (cf. the discussion in \cite{Minahan:2005mx}).
\item The one particle Green's function is not renormalized. Hence, the
dispersion relation (\ref{disper}) does not receive quantum
corrections.
\item The two-body S-matrix is given by the sum of bubble diagrams
(\figref{fig:LL-loops}).
\end{itemize}
These properties are almost obvious. A formal proof can be given by
cutting a generic diagram and counting intermediate propagators, as
illustrated in \figref{fig:LLcontour}.

\begin{figure}[t]
\begin{center}
\psfrag{p1}[c][c]{$p$} \psfrag{p2}[c][c]{$p'$}
\psfrag{k1}[c][c]{$k$} \psfrag{k2}[c][c]{$k'$}
\includegraphics*{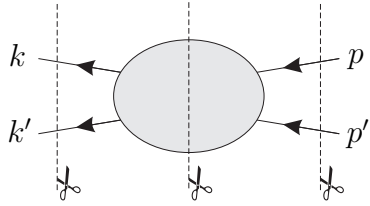}
\end{center}
\caption{\textbf{Cutting a generic diagram.} Due to charge
conservation the number of future directed propagators minus the
number of past directed propagators has to be the same at any moment
in time. Since any past directed propagator is identically equal to
zero, the number of propagators at any cut is the same and is equal
to the number of external incoming/outgoing legs.}
\label{fig:LLcontour}
\end{figure}

The fact that the two-body S-matrix is determined by the sum of bubble
diagrams in \figref{fig:LL-loops} has far reaching consequences.
Since these diagrams contain only quartic vertices we may truncate
the non-polynomial
action \eqref{LLscalar} at the fourth order in the fields%
\footnote{
The resulting Lagrangian is very similar to the one of the non-linear Schr\"{o}dinger model given by
$\Lagr = \frac{i}{2} \left(\pfi^* \partial_t \pfi - \partial_t \pfi^* \pfi \right) - \abs{\partial_x \pfi}^2 - g (\pfi^* \pfi)^2$.
For comparison we recall the S-matrix of this model: $S(p,p') = \frac{p - p' - ig}{p - p' + ig}$,
which can be computed by the same technique as we use here \cite{Thacker:1974kv}.}%
:
\begin{equation}\label{LLtru}
 \Lagr = \frac{i}{2} \left(\pfi^* \partial_t \pfi - \partial_t \pfi^* \pfi \right)
 -\abs{\partial_x \pfi}^2 - \frac{g}{2} \lrsbrk{ (\pfi^* \partial_x \pfi)^2 + (\partial _x \pfi^* \pfi)^2 }
 + \order{\pfi^6} \; .
\end{equation}
This is a very important simplification, which makes the all-loop
computation feasible and allows to forget about other non-linear
terms in the action. In \eqref{LLtru} we have introduced a formal
expansion parameter $g$ to make the power-counting of perturbative
series more transparent. We will set $g=1$ at the end of the
calculation. In fact, observable quantities cannot depend on $g$,
since this parameter can be eliminated by rescaling $t$, $x$ and
$\varphi$.

Let us now derive the Bethe equations for the  LL model
\eqref{LLaction} respectively \eqref{LLscalar}, i.e. compute the
two-particle S-matrix. The two relevant vertices, written in
\eqref{LLtru}, are
\begin{align}
- \frac{g}{2} \brk{\pfi^* \spa{\pfi}}^2 &
    \quad\longrightarrow\quad
    \psfrag{p1}[c][c]{$p$}
    \psfrag{p2}[c][c]{$p'$}
    \psfrag{k1}[c][c]{$k$}
    \psfrag{k2}[c][c]{$k'$}
    \raisebox{-6mm}{\includegraphics*{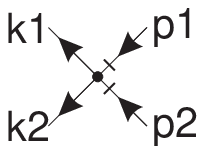}}
  = 2ig\, p_1 p'_1 \; , \\
- \frac{g}{2} \brk{\spa{\pfi^*} \pfi}^2 &
    \quad\longrightarrow\quad
    \psfrag{p1}[c][c]{$p$}
    \psfrag{p2}[c][c]{$p'$}
    \psfrag{k1}[c][c]{$k$}
    \psfrag{k2}[c][c]{$k'$}
    \raisebox{-6mm}{\includegraphics*{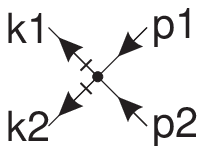}}
  = 2ig\, k_1 k'_1 \; .
\end{align}
The two-particle in- and out-states are defined as
\[
  \ket{p\, p'} = a^\dag(p) a^\dag(p') \ket{0}
  \comma
  \bra{k\, k'} = \bra{0} a(k') a(k)
  \; .
\]
The non-scattering part is given by
\[ \label{eqn:LL-non-scattering}
  \braket{k\, k'}{p\, p'} = \delta_+(p,p',k,k') \; ,
\]
where $\delta_+$ was defined in \eqref{eqn:emc}. At tree-level we need to evaluate the following expression
\[
  \left.\bra{k\, k'} \hat{S} \ket{p\, p'}\right|_{g}
  = \bra{k\, k'} \lrbrk{-\frac{ig}{2}} \int\!dt\,dx\, \bigsbrk{ \brk{\pfi^* \spa{\pfi}}^2 + \brk{\spa{\pfi^*} \pfi}^2 } \ket{p\, p'}
  \; .
\]
There are four identical Wick contractions between the states and each vertex leading to the factor $4(-kk'-pp')$. The integration over $t$ and $x$ imposes energy and momentum conservation, which can be written as
\[
 (2\pi)^2 \delta(p^2+{p'}^2-k^2-{k'}^2) \delta(p+p'-k-k')
 = \frac{1}{2(p-p')} \, \delta_+(p,p',k,k')
 \; .
\]
Multiplying all factors together we find
\[ \label{eqn:LL-S-one-loop}
  \left.\bra{k\, k'} \hat{S} \ket{p\, p'}\right|_{g}
  = 2\, ig \, \frac{pp'}{p-p'} \, \delta_+(p,p',k,k') \; .
\]
Including the non-scattering part \eqref{eqn:LL-non-scattering}, we can read off the S-matrix from comparison with \eqref{eqn:S-matrix} and find
to first order
\[
  S(p,p') = 1 + 2 \, ig \, \frac{pp'}{p-p'} + \order{g^2} \; .
\]

In \appref{app:LL-details} we compute all higher order corrections and find for arbitrary $n$, cf. \eqref{eqn:LL-all-loops}:
\[
  \left.\bra{k\, k'} \hat{S} \ket{p\, p'}\right|_{g^n}
  = 2 \, \lrbrk{ig}^n \, \lrbrk{\frac{p p'}{p-p'}}^n \, \delta_+(p,p',k,k') \; .
\]
Summing this result for all $n$ and adding the non-scattering part yields the exact S-matrix for the LL model
\[\label{LLS}
  S(p,p') = 1 + 2 \sum_{n=1}^\infty \lrbrk{ig}^n \lrbrk{\frac{p p'}{p-p'}}^n
          = \frac{\frac{1}{p}-\frac{1}{p'} - ig}
                 {\frac{1}{p}-\frac{1}{p'} + ig} \; ,
\]
where the expansion parameter $g$ can now be set equal to unity.

The Bethe equations resulting from this S-matrix are most
conveniently written in terms of the spectral variable $u=1/p$:
\begin{equation}\label{Lev}
e^{\frac{iL}{u_j}} = \prod_{k\neq j}^{}
             \frac{u_j-u_k + i}
                  {u_j-u_k - i} \; .
\end{equation}
The energy and the momentum are then given by
\[
 E = \sum_{j} \frac{\lambda}{8\pi^2u_j^2}
 \comma
 P=\sum_{j}^{}\frac{1}{u_j}
 \; .
\]

The Bethe equations for the spectrum of the Heisenberg Hamiltonian
\eqref{Werner} are very similar \cite{Faddeev:1996iy}. They are
obtained by replacing $1/u\rightarrow \pi -2\arctan u$ and
$1/u^2\rightarrow 1/(u^2+1/4)$ in the Bethe equations and the
dispersion relation above. The difference disappears at large $u$
(small momenta), as expected. In particular, the spectrum of the
low-energy modes with $p\sim 1/u\sim 1/L$, is the same for both
models and can be calculated from the approximate equation
\begin{equation}\label{low}
 \frac{L}{u_j}-2\pi n_j=\sum_{k\neq j}^{}\frac{2}{u_j-u_k}\,,
\end{equation}
up to and including finite-size  $\order{1/L}$ corrections
\cite{Minahan:2002ve}. As shown in \cite{Minahan:2005qj}, the
spectra for the two models start to deviate from one another at
order $\order{1/L^2}$.

The classical limit is achieved by assigning a macroscopic number of
Bethe roots to a finite set of mode numbers $n_I$. Then (\ref{low})
becomes an integral equation \cite{sutherland}:
\begin{equation}\label{class}
 \pint\frac{dy\,\rho (y)}{x-y}=\frac{1}{x}-2\pi n_I,
\end{equation}
where
\[
\rho (x)=\frac{1}{L}\sum_{j}^{}\delta \left(x-\frac{u_j}{L}\right) \; .
\]
It can be shown that (\ref{class}) describes all time-periodic
classical solutions of the LL equation (\ref{LLeqm})
\cite{Kazakov:2004qf}. This  classical limit of the Bethe equations
is obviously the same for the LL and the Heisenberg models.

The spectrum that follows from Bethe equations (\ref{Lev}) indicates
a rather unexpected instability. The bound states of elementary
excitations are described in integrable field theories by string
solutions of the Bethe equations. For instance, the 2-string
configuration is
\[
u_{1,2}=v\pm \frac{i}{2} \; .
\]
It becomes an exact solution of the Bethe equations in the strict
thermodynamic ($L\rightarrow \infty $) limit. The energy and the
momentum of the 2-string are:
\begin{eqnarray}\label{}
E_{{\rm 2-string}}&=&\frac{\lambda }{8\pi
^2}\left[\frac{1}{\left(v+\frac{i}{2}\right)^2}
+\frac{1}{\left(v-\frac{i}{2}\right)^2}\right]
 = \frac{\lambda }{4\pi ^2}\,\,
 \frac{v^2-\frac{1}{4}}{\left(v^2+\frac{1}{4}\right)^2} \; ,
\nonumber \\
P_{{\rm 2-string}}&=&\frac{1}{v+\frac{i}{2}}+\frac{1}{v-\frac{i}{2}}
=\frac{2v}{v^2+\frac{1}{4}} \; .
\end{eqnarray}
The momentum of the 2-string is always smaller that $2$, so the
2-string solutions of the Bethe equations describe two kinds of
excitations with the dispersion relations
\begin{equation}\label{}
 \varepsilon _\pm(p)=\frac{\lambda }{8\pi ^2}\left(\pm 2\sqrt{4-p^2}
 -4+p^2\right).
\end{equation}
The two branches arise from $v\gtrless 1/2$. The $\varepsilon _-$
branch has negative energy.

\begin{figure}[t]
\begin{center}
\psfrag{om}[l][l]{$\Omega$}
\psfrag{pm}[c][c]{$\varepsilon_-(P)$}
\psfrag{pp}[c][c]{$\varepsilon_+(P)$}
\psfrag{pq}[c][c]{$P^2/2$}
\includegraphics[width=8cm]{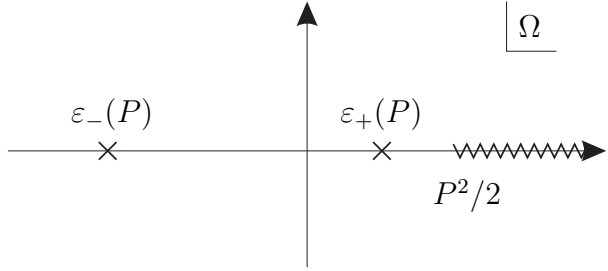}
\end{center}
\caption{\textbf{Analytic structure of the two-particle Green's
function \protect\eqref{2green}.}}
\label{fig:sp}
\end{figure}

The existence of negative-energy states seems to signal a vacuum
instability. Such an instability is absent in the Heisenberg spin
chain, and its appearance in the LL model is rather puzzling. We
have no simple explanation for this phenomenon, but we can
demonstrate that the negative-energy states do arise as poles of the
Green's functions. Consider, for instance, the Green's function of a
composite operator
\begin{equation}\label{defo}
 \mathcal{O}=\left(\varphi ^*\partial _x\varphi \right)^2
 +\left(\partial _x\varphi ^*\varphi \right)^2 \; ,
\end{equation}
which is computed in \appref{app:LL-instab}:
\begin{equation}\label{2green}
 \left\langle 0\right|T\,\mathcal{O}(t,x)\mathcal{O}(0,0)
 \left|\,0\right\rangle=
 \frac{i}{4}\int_{}^{}\frac{d\Omega \, dP}{(2\pi )^2}\,
 e^{-i\Omega t+iPx}\,
 \frac{(P^2-\Omega )^2\,}
 {P^2-\Omega -2\sqrt{P^2-2\Omega }} \; .
\end{equation}
As a function of $\Omega $, it has a two-particle cut from the
threshold at $\Omega =P^2/2$ to infinity and two poles at $\frac{\lambda}{8\pi^2}\Omega
=\varepsilon_\pm(P)$ (\figref{fig:sp}).


\section{Alday-Arutyunov-Frolov model}
\label{sec:aaf}

In this section we derive the Bethe equations for an integrable
system of two-dimensional fermions which was recently found by
Alday, Arutyunov and Frolov \cite{aaf:su11-string}. This model is a
new integrable quantum field theory whose Bethe-ansatz solution is
not known. Classical integrability of the model was demonstrated in
\cite{aaf:su11-string} by construction of a Lax representation for
the classical equations of motion. The model arises as a consistent
truncation of the classical $\AdS_5\times \Sphere^5$ superstring to
the $\algSU(1|1)$ subsector in the temporal gauge\footnote{In the
light-cone gauge the $\algSU(1|1)$ sector is described by free
world-sheet fermions \cite{af:su11-string}.}.


%
The AAF model is a theory of an interacting Dirac (two-component
complex) fermion in two dimensions%
\footnote{We use the Lagrangian
(5.7) from \cite{aaf:su11-string}, take the scaling (5.6) into
account and denote the angular momentum of the string by $L$ instead
of $J$. In addition we shifted $\sigma \to \sigma+2\pi
L/\sqrt{\lambda}$ to get positive limits for the integration.
Finally we drop the constant term in the action, which amounts to
shifting the origin of the energy scale.}:
\[ \label{eqn:aaf-model}
\begin{split}
\Action = \frac{\sqrt{\lambda}}{2\pi} \int\limits_{0}^{2\pi
L/\sqrt{\lambda}} \!\!\! d\sigma \int \! d\tau \, \Bigl[ &
          - \ihalf \bigbrk{ \bpsi \gamma^\alpha \partial_\alpha \psi - \partial_\alpha \bpsi \gamma^\alpha \psi }
          + \bpsi \psi \\[-3mm]
          &
          - \quarter \varepsilon^{\alpha\beta} ( \bpsi \partial_\alpha \psi \bpsi \gamma^3 \partial_\beta \psi
                                                  - \partial_\alpha \bpsi \psi \partial_\beta \bpsi \gamma^3 \psi ) \\[1mm]
          &
          + \sfrac{1}{8} \varepsilon^{\alpha\beta} ( \bpsi \psi )^2 \partial_\alpha \bpsi \partial_\beta \psi
          \Bigr] \; .
\end{split}
\]
The action is manifestly Lorentz-invariant. The Lorentz metric is
taken in the $(+1,-1)$ signature and the notation $\vx$ refers to
both components $(x^0,x^1) \equiv (\tau,\sigma)$. Below, we will use
the scalar product $\vec{a}\cdot\vec{b} = g^{\alpha\beta} a_\alpha
b_\beta$ and the ``vector'' product $\vec{a}\times\vec{b} =
\varepsilon^{\alpha\beta} a_\alpha b_\beta$, where the epsilon
tensor satisfies $\varepsilon^{01} = \varepsilon_{10} = +1$. The
explicit form of the Dirac matrices is given in
\eqref{eqn:Dirac-matrices} and the conjugate spinor is defined as
$\bpsi = \psi^\dag \gamma^0$.

In the action \eqref{eqn:aaf-model} all quantities ($\psi$,
$\sigma$, $\tau$, $L$, $\lambda$) are dimensionless. However, we
prefer to write it in a way which allows us to assign canonical mass
dimensions to the field and the coupling constants. Therefore we
rescale the world-sheet coordinates by $x^\alpha \to
\frac{\sqrt{\lambda}}{2\pi} x^\alpha$ and introduce the parameters
\[ \label{eqn:AAF-parameters}
  m = \frac{2\pi}{\sqrt{\lambda}}
  \comma
  g = \frac{\pi}{2\sqrt{\lambda}} \; .
\]
Then the space integral runs from 0 to $L$ and the Lagrangian can be written as%
\footnote{ Now this theory resembles the massive Thirring model:
$\Lagr = \bpsi ( i \slashed{\partial} - m ) \psi - \frac{g}{2}\,
\bpsi \gamma ^\alpha \psi\,\bpsi\gamma _\alpha \psi$. For later
comparison we recall the S-matrix of this theory: $S(\theta,\theta')
= \frac{1 - i \frac{g}{2} \tanh \frac{\theta - \theta'}{2}}{1 + i
\frac{g}{2} \tanh \frac{\theta - \theta'}{2}}$. We checked that this
S-matrix can be reproduced by the same diagrammatic calculation as
we do below for the AAF model.}
\[ \label{eqn:AAF-rewritten}
\begin{split}
\Lagr = & - \bpsi ( i \slashed{\partial} - m ) \psi \\
        & - \frac{g}{m^2} \varepsilon^{\alpha\beta} ( \bpsi \partial_\alpha \psi \bpsi \gamma^3 \partial_\beta \psi
                                                    - \partial_\alpha \bpsi \psi \partial_\beta \bpsi \gamma^3 \psi ) \\
        & + \frac{4g^2}{m^3} \varepsilon^{\alpha\beta} ( \bpsi \psi )^2 \partial_\alpha \bpsi \partial_\beta \psi \; .
\end{split}
\]
For the purpose of power counting, we may  assign the following mass
dimensions: $[\psi]=\half$, $[L]=[\vec{x}]=-1$, $[m]=1$, $[g]=0$.
Clearly, $m$ plays the role of a mass parameter and $g$ the role of
a coupling constant. The model is integrable for any $m$ and $g$. It
is not renormalizable by power-counting, but as conjectured in
\cite{aaf:su11-string}, the symmetries (integrability being one of
them) can take care of renormalization and render the model
renormalizable in perturbation theory, the same way symmetries make
two-dimensional non-linear sigma-models renormalizable. In this
sense, the AAF model is a fermionic counterpart of the non-linear
{\it sigma}-models.

One cannot compute the physical S-matrix of the AAF model directly
by resumming diagrams, as we did for the LL model, because the usual
relativistic propagator is not purely retarded and loop corrections
do not cancel. This difficulty is not fatal and can be overcome by
the trick that dates back to \cite{Berezin} and was used by
Bergknoff and Thacker to solve the massive Thirring model
\cite{Bergknoff:1978wr}. The idea is to use pseudo-vacuum (instead
of the true ground state) as a reference state for quantization.
The pseudo-vacuum, by definition, is a state annihilated by the
field operator:
\begin{equation}\label{}
 \psi(x) \ket{0} = 0 \; .
\end{equation}
All anti-particle levels in the pseudo-vacuum are left empty.
Quantizing in the pseudo-vacuum can thus be interpreted as applying
an infinite negative chemical potential to the system. The
scattering S-matrix of excitations over the pseudo-vacuum can be
computed exactly by essentially the same method as in any
non-relativistic theory. This ``bare'' S-matrix can be then used to
write down the Bethe equations. By solving the Bethe equations one
can fill back the Dirac sea and reconstruct the true ground state.
This is a pretty standard way to solve relativistic integrable field
theories \cite{Korepin_book}. In principle it allows one to derive
the spectrum of physical states and their S-matrix
\cite{Korepin:1979qq} from first principles. The price to pay is
that the solution of the Bethe equations that describes the ground
state and the solutions that describe physical excitations are
already rather complicated. It is obvious that the filling of the
Dirac sea drastically changes the spectrum of excitations and their
S-matrix.

By solving the free equations of motion we find the following mode
expansion of the fields
\begin{align} \label{eqn:AAF-mode-expansion}
  \psi(\vx)  & = \int \frac{dp_1}{2\pi} \, \lrsbrk{ a(p_1) u(p_1) \, e^{-i\vp\cdot\vx}
                                                  + b(-p_1) v(-p_1) \, e^{i\vp\cdot\vx} } \; , \\
  \bpsi(\vx) & = \int \frac{dp_1}{2\pi} \, \lrsbrk{ a^\dag(p_1) \bar{u}(p_1) \, e^{i\vp\cdot\vx}
                                                  + b^\dag(-p_1) \bar{v}(-p_1) \, e^{-i\vp\cdot\vx} } \; ,
\end{align}
where $p_0 = \sqrt{p_1^2 + m^2}$. The spinors $u(p_1)$ and $v(p_1)$
are defined in \appref{app:AAF-spinors}. The oscillators obey the
standard commutation relation:
\[
  \acomm{a(p_1)}{a^\dag(p'_1)} = 2\pi\,\delta(p_1-p'_1)
  \comma
  \acomm{b(p_1)}{b^\dag(p'_1)} = 2\pi\,\delta(p_1-p'_1)
  \; .
\]
The operators $a^\dag(p_1)$ and $b^\dag(p_1)$ create excitations
with momentum $p_1$ and with energies $+\sqrt{p_1^2 + m^2}$ and
$-\sqrt{p_1^2 + m^2}$.
We define the pseudo-vacuum $\ket{0}$ of the theory as the state
satisfying
\[ \label{eqn:pseudo-vacuum}
  a(p_1) \ket{0} = b(p_1) \ket{0} = 0 \qquad \mbox{for all $p_1$} \; .
\]
We call the excitations created by $a^\dag$ and $b^\dag$ on top of
this state pseudo-particles with positive and negative energy,
respectively.
The physical vacuum $\ket{\Omega}$ is obtained from $\ket{0}$ by
exciting all negative energy modes, i.e. by filling all holes in the
Dirac sea.

\begin{figure}[t]
\begin{center}
\psfrag{neg}[r][r]{$x^0 < {x'}^0$} \psfrag{pos}[r][r]{$x^0 >
{x'}^0$} \psfrag{kn}{$p_0$} \psfrag{km}[r][r]{$-\sqrt{p_1^2 + m^2} -
i\epsilon$} \psfrag{kp}[l][l]{$ \sqrt{p_1^2 + m^2} - i\epsilon$}
\includegraphics*{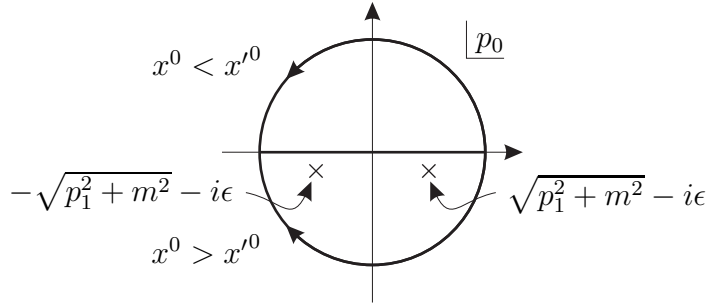}
\end{center}
\caption{\textbf{Pole prescription in Alday-Arutyunov-Frolov
model.}} \label{fig:AAFcontour}
\end{figure}

The major advantage of quantizing the theory in the pseudo-vacuum is
the absence of anti-particles, which implies the same
non-renormalization theorems as in the LL model of \secref{sec:ll}.
Since the field operator $\psi(\vx)$ annihilates the pseudo-vacuum,
the propagator is purely retarded:
\[ \label{eqn:AAF-propagator}
\begin{split}
 D(\vx-\vx')
 & := \bra{0} T \psi(\vx) \bpsi(\vx') \ket{0} \\
 & = (i\slashed{\partial} + m) \int\frac{d^2p}{(2\pi)^2} \, \frac{i}{\vp^2 - m^2 + i\epsilon(p_0)} \, e^{-i \vp\cdot(\vx-\vx')} \; ,
\end{split}
\]
where $\epsilon(p_0) = \sign(p_0) \epsilon$. The fact that
$D(\vx-\vx')$ vanishes for $x^0 < {x'}^0$ can easily be seen from
the pole prescription which is depicted in \figref{fig:AAFcontour}.
We mention that this pole prescription is relativistic invariant
because $\sign(p_0)$ is invariant under orthochronous Lorentz
transformations. Now, as both poles are in the lower half plane, we
conclude by precisely the same reasoning as in the previous section
that the closed loop of likewise oriented propagators vanishes, cf.
\figref{fig:nonrenorm}. The most important consequence of this
result is the fact that the two-body S-matrix is entirely determined
by the four-valent vertices alone. These are given by
\begin{align}
\label{eqn:AAF-vertex1} - \frac{g}{m^2} \varepsilon^{\alpha\beta}
\bpsi \partial_\alpha \psi \bpsi \gamma^3 \partial_\beta \psi &
    \quad\longrightarrow\quad
    \psfrag{p1}[c][c]{$p$}
    \psfrag{p2}[c][c]{$p'$}
    \psfrag{k1}[c][c]{$k$}
    \psfrag{k2}[c][c]{$k'$}
    \raisebox{-6mm}{\includegraphics*{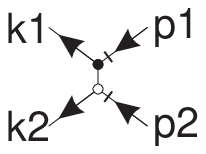}}
  = -\frac{ig}{m^2} \, \vpp \times \vp \; \unit \otimes \gamma^3
  \; , \\
\label{eqn:AAF-vertex2} + \frac{g}{m^2} \varepsilon^{\alpha\beta}
\partial_\alpha \bpsi \psi \partial_\beta \bpsi \gamma^3 \psi &
    \quad\longrightarrow\quad
    \psfrag{p1}[c][c]{$p$}
    \psfrag{p2}[c][c]{$p'$}
    \psfrag{k1}[c][c]{$k$}
    \psfrag{k2}[c][c]{$k'$}
    \raisebox{-6mm}{\includegraphics*{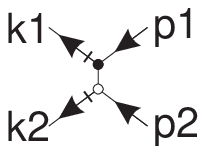}}
  = +\frac{ig}{m^2} \, \vkp \times \vk \; \unit \otimes \gamma^3
  \; .
\end{align}
The unit matrix $\unit$ (represented by the black dot) connects the
$p$ and $k$ legs, whereas $\gamma^3$ (represented by the white dot)
connects the $p'$ and $k'$ legs. The six-valent vertex is crucial
for the factorization of the S-matrix and hence for the
integrability of the theory, but it is of no relevance for the
scattering of two pseudo-particles!

\begin{figure}[t]
\begin{center}
\psfrag{p0}[c][c]{$p_0$} \psfrag{p1}[c][c]{$p_1$}
\psfrag{pos}[r][r]{\small $+\sqrt{p_1^2 + m^2}$}
\psfrag{neg}[r][r]{\small $-\sqrt{p_1^2 + m^2}$}
\psfrag{xia}[c][c]{\small $\theta \to - \infty$}
\psfrag{xib}[l][l]{\small $\theta = 0$} \psfrag{xic}[c][c]{\small
$\theta \to + \infty$} \psfrag{xid}[c][c]{\small $\theta \to i \pi
+\infty$} \psfrag{xie}[l][l]{\small $\theta = i \pi$}
\psfrag{xif}[c][c]{\small $\theta \to i \pi -\infty$}
\psfrag{pa}[cB][cB]{\small $( p_0,-p_1)$} \psfrag{pb}[cB][cB]{\small
$( p_0, p_1)$} \psfrag{pc}[cB][cB]{\small $(-p_0,-p_1)$}
\psfrag{pd}[cB][cB]{\small $(-p_0, p_1)$} \psfrag{zero}{\small $0$}
\psfrag{ipi}{\small $i\pi$} \psfrag{the}{$\theta$}
\includegraphics*{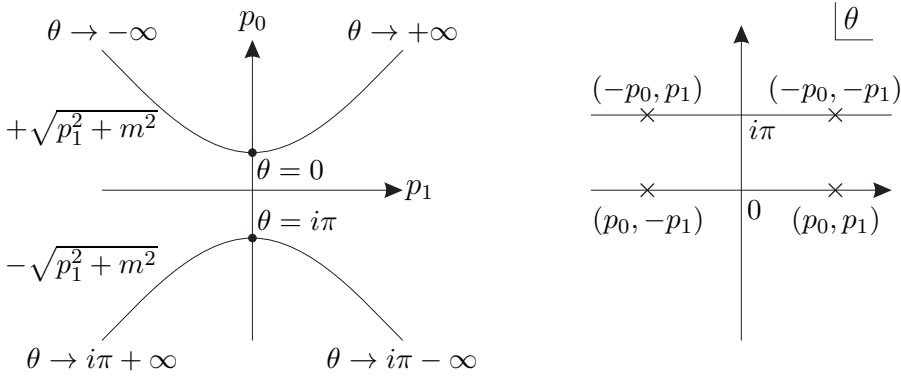}
\end{center}
\caption{\textbf{Rapidity plane.} It is convenient to work with a
complex rapidity $\theta$ as particles ($\Im\theta=0$) and
anti-particles ($\Im\theta=\pi$) can be treated at the same time.}
\label{fig:rapidity-plane}
\end{figure}

In the following we compute the S-matrix between two particles of
type $a^\dag$ (which have positive energy). The scattering of
particle of type $b^\dag$ can be obtained by analytic continuation
to complex rapidities $\theta$, defined by
\[ \label{eqn:def-rapidities}
  p_0 = m \cosh \theta
  \comma
  p_1 = m \sinh \theta
  \; .
\]
Real rapidities $\theta = \alpha \in \Reals$ parameterize the
mass-shell of the positive energy particles $a^\dag$, and complex
rapidities $\theta = i \pi - \alpha$ with $\alpha \in \Reals$
parameterize the mass-shell of the negative energy particles
$b^\dag$, cf. \figref{fig:rapidity-plane}.
As a meromorphic function of the rapidities, the S-matrix describes
the scattering of both types of particle at the same time. Hence it
is sufficient to consider the following in- and out-states:
\[
  \ket{p_1\, p'_1} = a^\dag(p_1) a^\dag(p'_1) \ket{0}
  \comma
  \bra{k_1\, k'_1} = \bra{0} a(k'_1) a(k_1)
  \; .
\]
The non-scattering part is given by
\[ \label{eqn:AAF-non-scattering}
  \braket{k_1\, k'_1}{p_1\, p'_1} = \delta_-(p_1,p'_1,k_1,k'_1) \; ,
\]
where $\delta_-$ was defined in \eqref{eqn:emc}. At tree-level we
need to evaluate the following expression
\[
  \left.\bra{k_1\, k'_1} \hat{S} \ket{p_1\, p'_1}\right|_{g}
  = \bra{k_1\, k'_1} \lrbrk{- \frac{ig}{m^2}} \int\!d^2x\, \varepsilon^{\alpha\beta}
    \bigsbrk{ \bpsi \partial_\alpha \psi \bpsi \gamma^3 \partial_\beta \psi
            - \partial_\alpha \bpsi \psi \partial_\beta \bpsi \gamma^3 \psi }
    \ket{p_1\, p'_1} \; .
\]
As the vertices do not have any symmetry, connecting the external
lines with the vertex leads to $2^3=8$ different terms. One of them
is given by
\[
  - \lrbrk{- \frac{ig}{m^2}}
    \brk{\vpp\times\vp} \;
    \bigbrk{\bar{u}(k'_1)\otimes\bar{u}(k_1)} \;
    \bigbrk{\unit\otimes\gamma^3} \;
    \bigbrk{u(p_1)\otimes u(p'_1)} \; .
\]
The others are obtained from this by exchange of the in-going
particles ($p \leftrightarrow p'$), by exchange of the out-going
particles ($k \leftrightarrow k'$) and by choosing the second vertex
($\vpp\times\vp \rightarrow \vkp\times\vk$). All changes are
accompanied by a change of the overall sign. The spacetime integral
imposes energy and momentum conservation, which can be written as
\[ \label{eqn:AAF-jacobian}
 (2\pi)^2 \delta(\vp+\vpp-\vk-\vkp) = \frac{p'_0 p_0}{\vpp\times\vp} \, \delta_+(p_1,p'_1,k_1,k'_1) \; .
\]
The $\delta$-functions in $\delta_+(p_1,p'_1,k_1,k'_1)$, cf.
\eqref{eqn:emc}, can be used to convert the $k$'s into $p$'s. After
this, the combination with the opposite relative sign
$\delta_-(p_1,p'_1,k_1,k'_1)$ forms. Eventually, the whole
expression boils down to
\[
\begin{split}
  \left.\bra{k_1\, k'_1} \hat{S} \ket{p_1\, p'_1}\right|_{g}
  = \frac{2ig}{m^2} & \,
    (\vpp\times\vp) \,
    \bigbrk{\bar{u}(p'_1)\otimes\bar{u}(p_1)} \,
    \bigbrk{\unit\otimes\gamma^3 - \gamma^3\otimes\unit} \,
    \bigbrk{u(p_1)\otimes u(p'_1)} \\
  & \cdot \frac{p'_0 p_0}{\vpp\times\vp} \, \delta_-(p_1,p'_1,k_1,k'_1) \; .
\end{split}
\]
The scalar product of the spinors is given by (for useful identities
see \appref{app:AAF-spinors})
\[
\begin{split}
  \bigbrk{\bar{u}(p'_1)\otimes\bar{u}(p_1)} \,
  \bigbrk{\unit\otimes\gamma^3 - \gamma^3\otimes\unit} \,
  \bigbrk{u(p'_1)\otimes u(p_1)} \,
  & = \frac{1}{2p'_0 p_0} \tr(\slashed{p}'+m)(\slashed{p}+m)\gamma^3 \\
  & = \frac{\vpp\times\vp}{p'_0 p_0} \; .
\end{split}
\]
Hence, at three level we find
\[
  \left.\bra{k_1\, k'_1} \hat{S} \ket{p_1\, p'_1}\right|_{g}
  = \frac{2ig}{m^2} \, (\vpp\times\vp) \, \delta_-(p_1,p'_1,k_1,k'_1) \; .
\]
Including the non-scattering part \eqref{eqn:AAF-non-scattering}, we
can read off the S-matrix as
\[
  S(p,p') = 1 + \frac{2ig}{m^2} \, (\vpp\times\vp) + \order{g^2} \; ,
\]
or in terms of rapidities \eqref{eqn:def-rapidities} as
\[
  S(\theta,\theta') = 1 + 2ig \sinh(\theta-\theta') + \order{g^2} \; .
\]

It is in fact possible to compute the S-matrix to all orders in $g$.
For this computation to be possible, it is essential to quantize the
theory in the pseudo-vacuum \eqref{eqn:pseudo-vacuum}. This is
because only in the pseudo-vacuum the higher loop diagrams are given
by the sum of bubble diagrams, cf. \figref{fig:AAF-loopdiagram}. The
full computation is nevertheless rather involved and therefore has
been relegated to \appref{app:AAF-higher-loops}. The higher order
corrections are given in \eqref{eqn:AAF-higher-loops} and lead
together with the non-scattering part to the exact S-matrix of the
AAF model:
\[ \label{ourS}
  S(p,p')
  = \frac{1+\frac{ig}{m^2} \, (\vpp\times\vp)}
         {1-\frac{ig}{m^2} \, (\vpp\times\vp)}
  \quad \mbox{or} \quad
  S(\theta,\theta')
  = \frac{1+ig \sinh (\theta-\theta') }
         {1-ig \sinh (\theta-\theta') } \; .
\]
Recall the value of the parameter $g$ and $m$ in terms of the 't
Hooft coupling $\lambda$ from \eqref{eqn:AAF-parameters}.

On the basis of integrability, the knowledge of the two-body
S-matrix is enough to immediately write down the Bethe equations:
\[ \label{eqn:AAF-BE}
  e^{i L m\sinh\theta_j} = \prod_{k\neq j}^{} \frac{1 - ig \sinh (\theta_j-\theta_k)}{1 + ig \sinh (\theta_j-\theta_k)} \; .
\]
Energy and momentum are given by
\[
  E = m \sum_{j} \cosh \theta_j
  \comma
  P = m \sum_{j} \sinh \theta_j
  \; .
\]

It is interesting to compare our result to the semiclassical
spectrum of the AdS string. Because of the fermionic nature of the
$\algSU(1|1)$ subsector, there are no classical spinning-string
solutions, but one can study states with very low momentum $p\sim
1/L$ (BMN states), which correspond to the lowest string modes.
$1/L$ corrections to the energies of these states were calculated
explicitly in  \cite{Callan:2003xr} and can be encoded in a compact
form in a set of Bethe equations \cite{Staudacher:2004tk} with the
S-matrix%
\footnote{To compare to \cite{Staudacher:2004tk} one has to
take into account that the R-charge denoted there by $J$ is what we
call $L$ here. If one identifies $L+({\rm number~of~Bethe~roots})/2$
with the length, as is done in \cite{Staudacher:2004tk}, then the
S-matrix effectively acquires an extra factor of $\exp[i(p'-p)/2]$,
cf. eq.~(4.31) in  \cite{Staudacher:2004tk}.}
\begin{equation}\label{Matthias}
  S(p,p')
  = \exp\left[\frac{i}{2}
  \left(p(\varepsilon(p')-\varepsilon(p)p'\right)
  \right] \; ,
\end{equation}
where%
\footnote{To be more precise, a lattice dispersion relation
with $p^2$ replaced by $4\sin^2(p/2)$ was postulated in
\cite{Staudacher:2004tk} . The difference, however, is $O(1/L^2)$
for the BMN states.}
\begin{equation}\label{redisp}
\varepsilon(p) = \sqrt{1 + \frac{\lambda}{4\pi^2} p^2} \; .
\end{equation}
The same Bethe equations were shown to arise from quantization of
the free fermions in the light-cone gauge \cite{af:su11-string}. We
now compare the S-matrix (\ref{Matthias}) to (\ref{ourS}). For
easier comparison, we give up our vector notation ($\vp$) and use
$p$ for the momentum and $\varepsilon$ for the energy. Furthermore
we have to take into account that we considered the AAF model with
rescaled time coordinate, i.e. we now compensate for this by
rescaling the energy by $\frac{\sqrt{\lambda}}{2\pi}$, so the
dispersion relation becomes precisely (\ref{redisp}) and the
S-matrix is
\[
  S(p,p')
  = \frac{1 + \frac{i}{4} \lrbrk{p\varepsilon(p')-\varepsilon(p)p'}}
         {1 - \frac{i}{4} \lrbrk{p\varepsilon(p')-\varepsilon(p)p'}}
  \; .
\]
It is immediately obvious that the S-matrices coincide in the low
momentum approximation, but deviate from each other at large
momenta.

We should warn the reader that the Bethe equations
(\ref{eqn:AAF-BE}) cannot be used to describe the quantum spectrum
of the string in $\AdS_5\times \Sphere^5$. Quantum corrections are
obviously different for the quantum field theory defined by
(\ref{eqn:AAF-rewritten}) and the full string sigma-model. It is
still interesting to note that quantum corrections in
(\ref{eqn:AAF-BE}) (deviations from  the classical phase shift
(\ref{Matthias})) are not of the form anticipated for the quantum
string in \cite{Arutyunov:2004vx,Beisert:2005cw}. We would also like
to stress that (\ref{eqn:AAF-BE}) are bare Bethe equations, they
have solutions with negative energies, and the true ground state is
the solution with the Dirac sea filled.

\begin{figure}[t]
\begin{center}
\psfrag{aj}[c][c]{$\alpha_j$}
\psfrag{ipi}{\small $i\pi$}
\psfrag{the}{$\theta$}
\includegraphics*{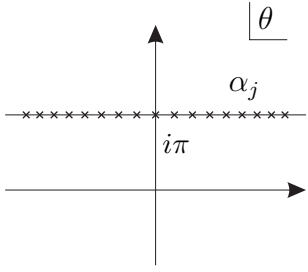}
\end{center}
\caption{\textbf{Bethe roots for physical vacuum.} With respect to
the pseudo-vacuum $\ket{0}$, which underlies the ``bare'' Bethe
equations, the physical vacuum $\ket{\Omega}$ is a highly excited
state. It corresponds to an $i\pi$-line completely filled with Bethe
roots $\alpha_j$.}
\label{fig:rapidity-plane-vacuum}
\end{figure}

We now turn to the discussion of the physical ground state of the
AAF model. Recall that for technical reasons we quantized the theory
in the pseudo vacuum $\ket{0}$, cf. \eqref{eqn:pseudo-vacuum}.
The physical vacuum $\ket{\Omega}$ is obtained from the
pseudo-vacuum by exciting all negative modes, symbolically
\[ \label{eqn:ket-omega}
  \ket{\Omega} = \prod_{p_1} b^\dag(p_1) \ket{0} \; .
\]
This state corresponds to a non-trivial solution of the bare Bethe
equations.
Recall that the $b^\dag$-particles are excitations with complex
rapidities $\theta_j = i\pi - \alpha_j$. Hence, the state
$\ket{\Omega}$  is described by placing Bethe roots as densely as
possible along the $i\pi$-line. This is depicted in
\figref{fig:rapidity-plane-vacuum}.
One can construct the vacuum solution of the Bethe equations in the
thermodynamic ($L\rightarrow \infty $) limit, find the spectrum of
physical excitations and their scattering S-matrix by solving
certain integral equations. The details of the derivation can be
found in the original literature on the massive Thirring model
\cite{Bergknoff:1978wr,Korepin:1979qq}, in the review
\cite{Thacker:1980ei} or in the monograph \cite{Korepin_book}. For
our model, the density of roots in the vacuum $\rho_\Omega(\alpha)$
satisfies the following equation
\[ \label{eqn:physical-vacuum-rho}
  m \cosh\alpha
  = 2\pi \rho_\Omega(\alpha)
  + \int_{-\infty }^{+\infty }\!d\bar\alpha\: \rho_\Omega(\bar\alpha) \,
  \frac{2g\cosh(\alpha-\bar\alpha)}{1+g^2\sinh^2(\alpha-\bar\alpha)}\,.
\]
The energy density of the physical vacuum is given by
\[ \label{eqn:physical-vacuum-energy}
  \frac{E}{L} = - m \int\!d\alpha\: \rho_\Omega(\alpha) \, \cosh\alpha \; .
\]
The physical excitations are obtained from the vacuum configuration
of \figref{fig:rapidity-plane-vacuum} by inserting further Bethe
roots outside the $i\pi$-line or by making holes in the Dirac sea.
The energy of those physical configurations is then measured with
respect to the ground state energy
\eqref{eqn:physical-vacuum-energy}.

\begin{figure}[t]
\begin{center}
\psfrag{mu}[l][l]{$\mu$}
\psfrag{a1}[c][c]{$-\sfrac{\pi}{2}$}
\psfrag{a2}[c][c]{$\sfrac{\pi}{2}$}
\psfrag{g1}[l][l]{$g\to0$}
\psfrag{g2}[l][l]{$g=1$}
\psfrag{g3}[l][l]{$g\to\infty$}
\psfrag{g4}[r][r]{$g\to-\infty$}
\psfrag{g5}[r][r]{$g=-1$}
\psfrag{g6}[r][r]{$g\to-0$}
\includegraphics*{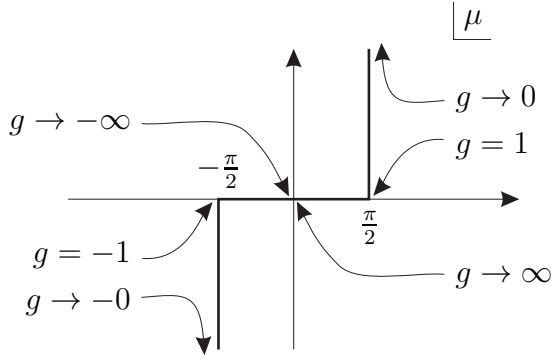}
\end{center}
\caption{\textbf{Coupling constant in AAF model.} It is convenient
to parameterize the coupling constant $g$ by a complex variable
$\mu$ in the way shown here. This is a solution of $\sin\mu=1/g$.}
\label{fig:AAF-coupling}
\end{figure}

Let us end this section with discussing the solutions of
\eqref{eqn:physical-vacuum-rho} in various regimes of the coupling
$g = \frac{\pi}{2\sqrt{\lambda}}$. In the attractive regime, $g<0$,
we find that $\rho_\Omega$ is determined only up to a constant. This
is because a constant shift of the density: $\rho_\Omega (\alpha
)\rightarrow \rho _\Omega (\alpha )+C$ does not affect the left hand
side of the equation. Choosing this constant arbitrarily large one
can make the energy \eqref{eqn:physical-vacuum-energy} arbitrarily
negative, which leads to a vacuum instability for any attractive
coupling constant. This behavior is similar to what happens in the
massive Thirring model at infinite coupling, which corresponds there
to the phase transition point. In the AAF model, however, the
instability is present for all values of $g<0$.

In the repulsive regime, $g>0$, there is no such problem.  We solve
\eqref{eqn:physical-vacuum-rho} by Fourier transformation. However,
we are required to introduce a rapidity cut-off
$\abs{\alpha}\le\Lambda$ in order to regularize the Fourier
transform of $\cosh\alpha$. Then the root density of $\ket{\Omega}$
can be written as the Fourier integral
\[ \label{eqn:rho-fourier-int}
  \rho_\Omega(\alpha)
  = \frac{m}{16\pi^2}
  \int_{-\infty}^\infty da\: e^{-i a \alpha} \,
  \frac{\cosh \frac{\pi}{2}a}{\cosh\frac{\mu}{2}a \, \cosh\frac{\pi-\mu}{2}a}
  \lrsbrk{\frac{e^{(1+ia)\Lambda}}{1+ia}
        + \frac{e^{(1-ia)\Lambda}}{1-ia}}
  \; .
\]
Here $\mu = \arcsin\brk{1/g}$, where we use the branches of
$\arcsin$ as depicted in \figref{fig:AAF-coupling}. Notice that
$\mu$ is real for strong coupling $g>1$, but complex for weak
coupling $0<g\le1$. The integral \eqref{eqn:rho-fourier-int} can be
computed by contour integration. In the limit $\Lambda\to\infty$,
the dominant contribution to the integral stems from the poles at $a
= \pm i \frac{\pi}{\pi-\mu}$ in the strong coupling regime and from
the poles at $a = \pm i \frac{\pi}{\pi-\mu}$ and $a = \pm i
\frac{\pi}{\pi-\mu^*}$ in the weak coupling regime. For $g>1$ we
have
\[ \label{eqn:rho-solution}
  \rho_\Omega(\alpha)
  = \frac{m}{2\pi} \,
  e^{-\sfrac{\pi}{\pi-\mu}\Lambda} \;
  \frac{1}{\mu} \,
  \cot\frac{\pi^2}{2(\mu-\pi)} \,
  \cosh\frac{\pi\alpha}{\pi-\mu} \; ,
\]
for $0<g\le1$  we have \eqref{eqn:rho-solution} plus it complex
conjugate. In the strong coupling regime, we can introduce a
renormalized mass
\[ \label{eqn:renormalized-mass}
  m_R \sim m \, e^{-\sfrac{\pi}{\pi-\mu}\Lambda}
\]
in order to get  a finite result when the cut-off is removed. In the
weak coupling regime this is not possible as the two terms in
$\rho_\Omega(\alpha)$ have a different dependence on $\Lambda$.
Hence, the model is non-renormalizable for $0<g =
\frac{\pi}{2\sqrt{\lambda}}\le1$. It is instructive to express
\eqref{eqn:renormalized-mass} in terms of the original coupling $g$.
In the limit $g\to0$, we find that the anomalous dimension of the
mass $m$ behaves as
\[
  \gamma _m = \frac{\pi}{\pi-\mu} \sim \frac{i}{\ln g} \; .
\]
The approach of the anomalous dimension to zero as $g\rightarrow 0$
is not analytic. We see that the breakdown of renormalizability is a
non-perturbative effect.

In the strong repulsive regime ($g>1$), the AAF model seems to be a
well defined quantum theory.
Unfortunately, additional complications arise in this case, because
one can further diminish the vacuum energy by inserting $n$-strings
around the $i\pi$-line. We thus expect that the true vacuum is a
condensate of $n$-strings, where $n$ can depend on $g$, like in the
strongly repulsive Thirring model \cite{Korepin:1979hg}.

\section{Faddeev-Reshetikhin model}
\label{sec:fr}

In this section we study string theory on $\Sphere^3\times \Reals$.
We should probably explain what we mean by that because
$\Sphere^3\times \Reals$ is not a string background. There are no
problems with classical strings, in fact $\Sphere^3\times \Reals$
can be regarded as a subspace of $\AdS_5\times \Sphere^5$, but
quantization leads to UV divergences and non-zero beta-function.
Although the resulting model cannot be interpreted as string theory,
it is an interesting example of two-dimensional integrable field
theory first considered in \cite{fr:principle-chiral-field}.

The string action in the conformal gauge is\footnote{The string
tension of the AdS string is $\sqrt{\lambda }/2\pi $.}
\begin{equation}\label{sigmaaction}
 S=-\frac{\sqrt{\lambda }}{4\pi }\int_{}^{}d\tau d\sigma \,
 \left[\frac{1}{2}\,\tr j_a^2+\left(\partial
 _aX^0\right)^2\right],
\end{equation}
where $X^0$ is the time coordinate and
\[
j_a=g^{-1}\partial _a g \; .
\]
Here $g$ is a group element of $SU(2)$ that parameterizes an
embedding of the string world-sheet in $\Sphere^3$. The current
satisfies the following equations of motion:
\begin{eqnarray}\label{}
\partial _aj^a&=&0 \; , \nonumber \\
\partial _aj_b-\partial _bj_a+[j_a,j_b]&=&0 \; .
\end{eqnarray}
In addition we should impose the Virasoro constraints:
\begin{equation}\label{}
 \tr j_\pm^2=-2\left(\partial _\pm X^0\right)^2 \; ,
\end{equation}
where $\partial_\pm = \partial _\tau \pm\partial _\sigma $ are
derivatives with respect to the light-cone coordinates $\sigma
^\pm=(\tau \pm\sigma )/2$. Accordingly, $j_\pm$ are the light-cone
components of the $su(2)$ current.

The standard way to proceed would be to quantize the sigma-model and
then impose the Virasoro constraints in the weak sense -- as
subsidiary conditions on the physical states. This procedure is
inconsistent for the model at hand, for the reasons explained above.
It is also not likely to work for the full string sigma-model in
$\AdS_5\times \Sphere^5$ because of the well-known problems with the
conformal gauge for the Green-Schwarz superstring. Another (in fact,
no less standard) approach to string quantization is to fix the
gauge completely, solve the Virasoro constraints (eliminate
transverse degrees of freedom), and only then quantize. We thus
choose the temporal gauge:
\begin{equation}\label{}
 X^0=\kappa \tau \; ,
\end{equation}
such that the Virasoro constraints become
\begin{equation}\label{}
\,\tr j_\pm^2=-2\kappa ^2 \; ,
\end{equation}
and are solved by
\begin{equation}\label{}
 j_\pm=i\kappa \vec{S}_\pm\cdot\vec{\sigma } \; ,
\end{equation}
where $\vec{S}_+$ and $\vec{S}_-$ are three-dimensional vectors of
unit norm: $\vec{S}_\pm^2=1$. The satisfy the equations of motion
\begin{eqnarray}\label{equa}
 \partial _+S_-^i+\kappa \varepsilon ^{ijk}S_-^jS_+^k&=&0 \; ,\nonumber \\
 \partial _-S_+^i+\kappa \varepsilon ^{ijk}S_-^jS_+^k&=&0 \; .
\end{eqnarray}

Now we face an immediate problem: What Hamiltonian should be quantized? The
original Hamiltonian, canonical conjugate to $\tau $ in
(\ref{sigmaaction}) is set to zero by the Virasoro constraints. The
way out is to use another Hamiltonian and another Poisson structure,
proposed in \cite{fr:principle-chiral-field}. They are more natural
from the point of view of integrability, and are potentially
relevant in the AdS/CFT correspondence
\cite{Zarembo:2004hp,Mikhailov:2005sy}. The action in the path
integral for the evolution operator associated with the Hamiltonian
of \cite{fr:principle-chiral-field} is
\begin{equation}\label{}
 S=\int_{}^{}d^2x\,\left[
 C_+(\vec{S}_-)+C_{-}(\vec{S}_+)-\frac{\kappa}{2}\, \vec{S}_+\cdot
 \vec{S}_-
 \right] \; ,
\end{equation}
where $C_\pm$ are Wess-Zumino terms as defined in \eqref{eqn:non-local-WZ}.
This will be our starting point. It is easy to check that the
equations of motion (\ref{equa}) follow from variation of this
action.

To develop perturbation theory for the FR model, we will perform the
same change of variables as in sec.~\ref{sec:ll}:
\begin{equation}\label{}
 \phi _\pm=\frac{S_\pm^1+iS_\pm^2}{\sqrt{2+2S_\pm^3}} \comma
 \qquad S^3_\pm=1-2|\phi _\pm|^2 \; .
\end{equation}
Upon this change of variables, the Wess-Zumino term becomes a
canonically normalized first-order action for $\phi _+$, $\phi _-$:
\begin{eqnarray}\label{ddir}
 S&=&\int_{}^{}d^2x\,\left[
 \frac{i}{2}\left(\phi _+^*\partial _-\phi _+-\partial _-\phi _+^*\phi _+
 +\phi _-^*\partial _+\phi _--\partial _+\phi _-^*\phi _-
 \right)
 \right. \nonumber \\ &&\left.\vphantom{\frac{i}{2}}
 -\kappa\sqrt{\left(1-|\phi _+|^2\right)\left(1-|\phi _-|^2\right)}
 \left(\phi _+^*\phi _-+\phi _-^*\phi _+\right)
 +\kappa \left(|\phi _+|^2+|\phi _-|^2\right)
 \right. \nonumber \\ &&\left.\vphantom{\frac{i}{2}}
 -2\kappa |\phi _+|^2|\phi _-|^2
 \right] \; .
\end{eqnarray}
This action can be cast into a very concise form if we combine $\phi
_+$ and $\phi _-$  into a two-component commuting  spinor:
\begin{equation}\label{}
 \phi =
 \left(\begin{array}{c}
     \phi _-  \\
      \phi _+ \\
 \end{array}\right) \; .
\end{equation}
Then (\ref{ddir}) becomes a Dirac-like action%
\footnote{To make power-counting possible we introduce the mass
$m=\kappa$ and the coupling constant $g=\kappa/2$.}
\begin{equation}\label{Paul}
 S=\int_{}^{}d^2x\,\left(
 i\bar{\phi }\not{}\!\!D\phi -m\bar{\phi }\phi -g\,\bar{\phi }\gamma ^\mu \phi
 \,\,\bar{\phi }\gamma _\mu \phi +O(\phi ^6)
 \right) \; .
\end{equation}
The covariant derivative contains a field-dependent chemical
potential:
\begin{equation}\label{covar}
 D_0 = \partial_0-im-ig\,\bar{\phi}\phi \comma
 D_1 = \partial_1 \; .
\end{equation}

This action describes a single charged particle and its
anti-particle with the dispersion relations
\begin{equation}\label{BMN1}
 \varepsilon = \sqrt{p^2+m^2}-m\qquad \mbox{(particle)} \; ,
\end{equation}
and
\begin{equation}\label{BMN2}
 \varepsilon =\sqrt{p^2+m^2}+m\qquad \mbox{(anti-particle)} \; .
\end{equation}
This first of these equations is the BMN formula
\cite{Berenstein:2002jq}. The mass gap for the particles is offset
to zero by the chemical potential. The energy of anti-particles is
shifted in the opposite direction, so that anti-particles decouple
at low energies and momenta. The low-energy effective theory can be
constructed by defining big and small components of the spinor:
\begin{equation}\label{}
 \pfi =\frac{1+\gamma ^0}{2}\,\phi ,\qquad \chi =\frac{1-\gamma
 ^0}{2}\,\phi \; .
\end{equation}
The Lagrangian  in (\ref{Paul}) takes the form
\begin{equation}\label{}
 \mathcal{L}=i\,\pfi ^*\partial _0\pfi +i\,\chi^* \partial _0\chi
 +i\,\pfi ^*\partial _1\chi +i\,\chi ^*\partial _1\pfi +2m|\chi |^2
 +g\left({\pfi ^*}^2\chi ^2+{\chi ^*}^2\pfi -|\chi
 |^4\right) \; .
\end{equation}
Integrating out $\chi $ we arrive at
\begin{equation}\label{}
 \mathcal{L}=i\,\pfi ^*\partial _0\pfi-\frac{1}{2m}\,|\partial _1\pfi
 |^2-\frac{g}{4m^2}\left[\left({\pfi^*}\partial _1\pfi \right)^2
 +\left(\partial _1\pfi ^*\pfi \right)^2 \right]+\ldots \; ,
\end{equation}
which is the same as (\ref{LLtru}). This is another way to see that
the Landau-Lifshitz model is the low-energy effective theory for
strings on $\Sphere^3\times \Reals$.

Returning to the sigma-model, we can now quantize (\ref{Paul}) in
the ``wrong'' vacuum, in which all anti-particle states are empty.
Then the S-matrix can be computed  by summing the bubble diagrams.
If the pole prescription is non-relativistic, the field-independent
part of the chemical potential in (\ref{covar}) can be eliminated by
a shift of the integration variable $k^0$, and we can use the
relativistic dispersion relation instead of the BMN formulas
(\ref{BMN1}), (\ref{BMN2}). Calculation of the loop integral for the
$pp'\rightarrow pp'$ scattering (the details can be found in
appendix~\ref{loopinFR}) gives:
\begin{equation}\label{loopFR}
 {\rm Loop}=\frac{(\slashed{p}'+m)\otimes(\slashed{p}+m)
 +(\slashed{p}+m)\otimes(\slashed{p}'+m)}
 {8m^2\sinh(\theta -\theta ')}\;.
\end{equation}
The S-matrix is%
\footnote{We change our conventions for the wave functions of
external states. Here
$$u(p)= \sqrt{m}\left(\begin{array}{c}
    e^{-\theta /2} \\
    e^{\theta /2}  \\
\end{array}\right)$$
describes pseudo-particles with both positive and negative
energies. The spinors are normalized as $\bar{u}(p)u(p)=2m$.}
\begin{eqnarray}\label{}
 \bra{k\, k'} \hat{S} \ket{p\, p'}
 &=&4p_0p_0'(2\pi )^2\delta _+(k,k',p,p')
 \nonumber \\ &&
  +(2\pi )^2\delta ^{(2)}(k+k'-p-p')\,\frac{ig}{2}\,\bra{\rm out}
 \left(\gamma ^0\otimes 1+1\otimes \gamma ^0-2\gamma ^\mu \otimes\gamma _\mu \right)
 \nonumber \\ && \nonumber \times
 \sum_{n=0}^{\infty }\left[\frac{\slashed{p}'+m)\otimes(\slashed{p}+m)
 +(\slashed{p}+m)\otimes(\slashed{p}'+m)}
 {8m^2\sinh(\theta -\theta ')} \right. \nonumber \\
 && \qquad\qquad \times \left.
 \vphantom{\frac{(\slashed{p}'+m)\otimes(\slashed{p}+m)
 +(\slashed{p}+m)\otimes(\slashed{p}'+m)}
 {8m^2\sinh(\theta -\theta ')} }
 ig
 \left(\gamma ^0\otimes 1+1\otimes \gamma ^0-2\gamma ^\alpha \otimes\gamma _\alpha \right)
 \right]^n
 \ket{\rm in},
\end{eqnarray}
where by $\ket{\rm in}$ and $\bra{\rm out}$ we denote bi-spinors
\begin{eqnarray}\label{}
 \ket{\rm in} & = & u(p)\otimes u(p')+u(p')\otimes u(p) \; , \nonumber \\
 \bra{\rm out}& = & \bar{u}(p)\otimes \bar{u}(p') + \bar{u}(p')\otimes \bar{u}(p) \; .
\end{eqnarray}
Observing that
\[
\slashed{p}+m=u(p)\bar{u}(p) \; ,
\]
and taking into account the symmetry of the vertex, we can replace
\[
 (\slashed{p}'+m)\otimes(\slashed{p}+m)
 +(\slashed{p}+m)\otimes(\slashed{p}'+m)
 \rightarrow
 \frac{1}{2} \ket{\rm in}\bra{\rm out} \; .
\]
Then the S-matrix becomes
\begin{equation}\label{}
 S(\theta ,\theta ')=1+2\sum_{n=1}^{\infty }
 \left[\frac{ig\left\langle {\rm out}\right|
 \left(\gamma ^0\otimes 1+1\otimes \gamma ^0-2\gamma ^\alpha \otimes\gamma _\alpha \right)
 \left|\,{\rm in}\right\rangle}{16m^2\sinh(\theta -\theta ')}\right]^n \; .
\end{equation}
Plugging the explicit form of the wave functions $\bar{u}(p)$ and
$u(p)$ into the matrix element of the vertex, we finally get:
\[
 S(\theta,\theta')
 = \frac{1+ig \left( \frac{\cosh \frac{\theta+\theta'}{2}}{\sinh \frac{\theta-\theta'}{2}} - \coth \frac{\theta-\theta'}{2} \right) }
        {1-ig \left( \frac{\cosh \frac{\theta+\theta'}{2}}{\sinh \frac{\theta-\theta'}{2}} - \coth \frac{\theta-\theta'}{2} \right) }
 \; .
\]
It is easy to check that in the limit of small momenta the S-matrix
of the LL model \eqref{LLS} is recovered.

The S-matrix of the FR model is not Lorentz invariant. This is not
surprising because the Virasoro constraints explicitly break the
Lorentz symmetry of the originally Lorentz-invariant chiral field.
Because of the lack of Lorentz invariance the S-matrix is not a
function of $\theta -\theta '$, which makes rapidity
parameterization not the most convenient one. There is another
parameterization of energies and momenta which is much more useful
in this particular case:
\begin{eqnarray}\label{}
 &&\frac{\varepsilon }{m}=\cosh\theta =\frac{x^2+1}{x^2-1}\,,\nonumber \\
 &&\frac{p}{m}=\sinh\theta =\frac{2x}{x^2-1} \; .
\end{eqnarray}
The S-matrix takes an extremely simple form in these variables:
\begin{equation}\label{}
 S(x,x')=\frac{x-x'-2ig}{x-x'+2ig} \; .
\end{equation}

We can now write down the Bethe equations:
\begin{equation}\label{besu2}
 \exp\left({\frac{2imLx_j}{x^2_j-1}}\right)=\prod_{k\neq j}^{}
 \frac{x_j-x_k+2ig}{x_j-x_k-2ig} \; .
\end{equation}
The states with Bethe roots in the interval $-1<x_j<1$ carry
negative energy. In quantum theory all of the negative-energy levels
should be filled. The repulsive ($g<0$) and the attractive ($g>0$)
cases are very different in this respect.  In the repulsive case,
the roots cannot form strings and the ground-state density in the
thermodynamic limit is determined by the following integral equation
\begin{equation}\label{}
 m\,\frac{1+x^2}{(1-x^2)^2}=\pi \rho (x)
 -2|g|\int_{-1}^{1}\frac{dy\,\rho (y)}{(x-y)^2+4g^2} \; .
\end{equation}
The singularities at $x=\pm 1$ require regularization and will lead
to renormalization of the bare parameters. It would be interesting
to solve this equation and to perform the renormalization
explicitly.

The situation is more complicated in the attractive regime of $g>0$.
In this case, both positive- and negative-energy roots can form
strings with $\Delta x=2ig$. A simple calculation shows that the
energy  of the $n$-string with the centre of mass between $-1$ and
$1$ decreases with $n$ (becomes more and more negative) up to $n\sim
1/2g$. We thus expect that the ground state is a condensate of
$n$-strings with very large $n\sim 1/g$. This is in a qualitative
agreement with the results of Faddeev and Reshetikhin for the
lattice-regularized model \cite{fr:principle-chiral-field}. The
vacuum of the lattice model is a condensate of $2S$-strings with
$S\rightarrow \infty $. We will not attempt to solve the Bethe
equations (\ref{besu2}) here, but we expect that the renormalized
solution describes the quantized principal chiral field
\cite{Polyakov:1983tt}, as it is the case for the lattice model of
\cite{fr:principle-chiral-field}.

\section{Conclusions}

We hope that our calculations illustrate several features of the
Bethe ansatz that can be useful in solving string theory in
$\AdS_5\times \Sphere^5$. The consistency of the approach we used fully
relies on quantum integrability which reduces the many-particle
problem to superposition of two-body interactions. This allowed us
to avoid complexities associated with highly non-linear interactions
 in the sigma-models. Let us take the LL model as an example.
Its Lagrangian contains infinitely many non-linear terms, but we
needed only the quartic vertices to derive the Bethe equations and
thus to reconstruct the full spectrum. In other words we managed to
solve the model by analyzing just the small fluctuations around the
"north pole" $\vec{n}=(0,0,1)$. To illustrate this point, imagine
that the target space of the LL sigma-model (the sphere $\Sphere^2$)
is deformed near the south pole, for instance that the constraint
$\vec{n}^2=1$ is replaced by $\vec{n}^2=f(n_3)$. If $f(1-\varepsilon
)=1+\order{\varepsilon^3}$, the quartic vertices of perturbation
theory around $\vec{n}=(0,0,1)$ are still the same and we will get
the same two-body S-matrix. The complete non-perturbative spectrum
of the deformed model, however, is totally different. The full
spectrum is certainly not determined by the two-body S-matrix,
because the deformation completely destroys the integrability and
multi-particle interactions do not factorize any more.

To further illustrate the power of integrability, we note that the
Bethe ansatz completely determines the spectrum of a system in a
finite volume, which is particularly important for applications to
string theory, but we did not use periodic boundary conditions
anywhere in deriving Bethe equations. In fact, the S-matrix is only
defined in the infinite volume where the notion of asymptotic states
makes sense.

Although Bethe ansatz completely determines the spectrum, it may
happen that physically interesting states with low energies are
described by rather complicated solutions of the Bethe equations.
The ground state for a relativistic system typically is a Dirac sea
that contains a non-trivial distribution of infinitely many
roots\footnote{We would like to thank F.~Smirnov for the discussion
of this point.}. Usually one can find the ground-state distribution
in a closed form only in the thermodynamic limit. It is interesting
to note here that anti-particle-like states with negative energies
seem to also arise on the gauge-theory side of the AdS/CFT
correspondence. The all-loop Bethe equations for the spectrum of
local operators in $\mathcal{N}=4$ SYM \cite{Beisert:2005fw} do have
negative-energy solutions at the non-perturbative level
\cite{Beisert:2005tm,Rej:2005qt}. An interpretation and a physical
significance of these states remains unclear to us. In particular,
in \cite{Rej:2005qt} the negative-energy states were projected out
rather than filled. We believe that these states are direct
counterparts of the anti-particles in the sigma-model. Understanding
their role in the AdS/CFT correspondence is a very interesting and
important problem.

\paragraph{Note added.} A complementary approach to the quantum
Bethe ansatz in the string-like sigma-models, which is
 based on the known
exact physical S-matrices \cite{zz:factorized-s-matrix}, was
developed in a parallel publication \cite{4a}.

\subsection*{Acknowledgments}

We would like to thank N.~Gromov, V.~Kaza\-kov, I.~Kostov, K.~Sakai,
D.~Serban, F.~Smirnov, M.~Staudacher, A.~Tseyt\-lin and P.~Vieira
for illuminating discussions. K.Z. would like to thank Laboratoire
de Physique Th\'eorique de l'Ecole Normale Sup\'erieure, where this
work was completed, for kind hospitality. This work was supported in
part by the G\"oran Gustafsson Foundation. The work of K.Z. was also
supported in part by the Swedish Research Foundation under contracts
621-2004-3178 and 621-2003-2742, and by the RFBR grant
NSh-1999.2003.2 for the support of scientific schools.

\appendix

\section{Computational details for LL model}
\label{app:LL-details}

\subsection{Loop integrals}
\label{app:LL-loop-integrals}

A generic higher loop diagram in the LL model consists of a chain of bubbles as depicted in \figref{fig:LL-loops}. In this appendix we compute the ``bubble propagators'' $I_r$. More specifically, we compute the momentum space representation of two parallel propagators $D(t,x)$, cf. \eqref{eqn:LL-propagator}, with two inflowing on-shell momenta $p$ and $p'$ ($p>p'$). We need to consider the cases with $r=0,1,2$ pairs of derivatives acting onto the two propagators. In the case without derivatives we have
\[
\begin{split}
I_0(p,p') & = \psfrag{p1}[c][c]{$p$}
              \psfrag{p2}[c][c]{$p'$}
              \psfrag{label1}[c][c]{$q$}
              \psfrag{label2}[c][c]{$p+p'-q$}
              \raisebox{-9mm}{\includegraphics*{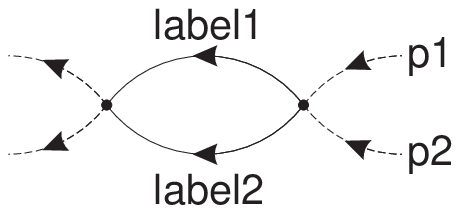}} \\[3mm]
          & = \int\!dt\,dx\, \lrbrk{D(x,t)}^2 \; e^{i(p^2+{p'}^2)t - i(p+p')x} \\
          & = \int\frac{d\omega\, dq}{(2\pi)^2} \,
              \frac{i^2}{\bigsbrk{\omega - q^2 + i\epsilon}
                         \bigsbrk{(p^2+{p'}^2-\omega) - (p+p'-q)^2 + i\epsilon}}
          \; .
\end{split}
\]
First perform the energy integral over $\omega$ by contour integration. The integrand has one pole in the upper and one pole in the lower half plane. Closing the contour in either half plane leads to
\[
I_0(p,p') = -\frac{i}{2} \int\frac{dq}{2\pi} \,
             \frac{1}{(q-\half p -\half p')^2 - \quarter (p-p')^2 - i\epsilon} \; .
\]
Shift the integration variable by $q\to q+\half p+\half p'$ and perform the momentum integration also by contour integration. One finds
\begin{equation}\label{loop0}
I_0(p,p') = \frac{1}{2 (p-p')} \; .
\end{equation}

Loops with derivatives, see \eqref{eqn:LL-I1} and \eqref{eqn:LL-I2}, are similar to the above. Every pair of derivatives introduces a factor of $-k(p-k)$ into the numerator. The energy integral in unchanged. However, the momentum integral is now divergent. For the evaluation of these integrals we invoke dimensional regularization. This has the same effect as adding appropriate counterterms to the action. The integrals become finite and their values are
\begin{eqnarray}
 I_1(p,p') &=& \frac{-pp'}{2 (p-p')}\,,
 \\ \label{loop2}
 I_2(p,p') &=& \frac{(pp')^2}{2 (p-p')}\,.
\end{eqnarray}

\subsection{Summing all Feynman diagrams}

We compute
\[
  \left.\bra{k\, k'} \hat{S} \ket{p\, p'}\right|_{g^n}
  = \bra{k\, k'} \, \frac{1}{n!} \lrsbrk{ \lrbrk{-\frac{ig}{2}} \int\!dx\,dt\, \bigsbrk{ \brk{\pfi^* \spa{\pfi}}^2 + \brk{\spa{\pfi^*} \pfi}^2 } }^n \ket{p\, p'} \; .
\]
The vertices are to be connected as in \figref{fig:LL-loops}. Any other diagram will be zero. There are $2^{n+1} n!$ different ways of connecting $n$ four-vertices among themselves and to the external legs in the way of \figref{fig:LL-loops}. Furthermore we need to take into account that we have two different vertices. This amounts to placing derivatives onto the legs of the vertex, either onto the pair of in-going or on the pair of out-going lines. This leads to $2^n$ different diagrams.

A generic diagram consists of a chain of loops. If the derivatives hit the external legs we get a factor of $-pp'$. Derivatives on internal lines lead to three different kinds of loops:
\begin{align}
I_0(p,p') & = \psfrag{p1}[c][c]{$p$}\psfrag{p2}[c][c]{$p'$}\raisebox{-6mm}{\includegraphics*{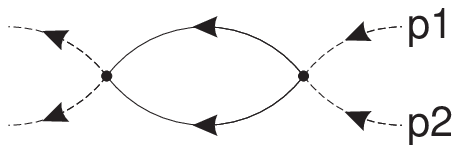}} \\
I_1(p,p') & = \psfrag{p1}[c][c]{$p$}\psfrag{p2}[c][c]{$p'$} \raisebox{-6mm}{\includegraphics*{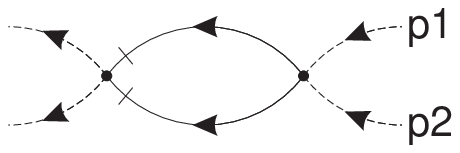}}
            =                                               \raisebox{-6mm}{\includegraphics*{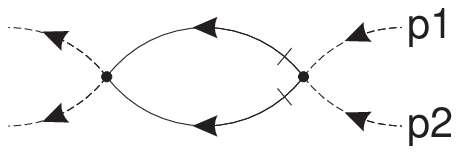}} \label{eqn:LL-I1} \\
I_2(p,p') & = \psfrag{p1}[c][c]{$p$}\psfrag{p2}[c][c]{$p'$} \raisebox{-6mm}{\includegraphics*{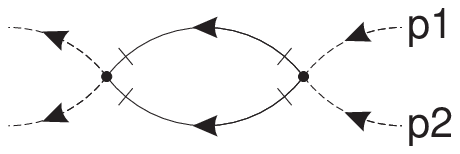}}  \label{eqn:LL-I2}
\end{align}
The corresponding loop integrals are evaluated in the previous subsection. Counting how many of the $2^n$ diagrams lead to a given expression
\[
 (-p p')^{\#_1} I_0^{\#_2}(p,p') I_1^{\#_3}(p,p') I_2^{\#_4}(p,p')
\]
is not completely trivial in general. This counting, however, becomes unnecessary as we find
\[
  I_1(p,p') = (-p p')   I_0(p,p')
  \comma
  I_2(p,p') = (-p p')^2 I_0(p,p')
  \; .
\]
This means that every diagram, no matter how the derivatives are distributed, will contribute
\[
 (-p p')^n \, I_0^{n-1}(p,p') = \frac{(-p p')^n}{\bigbrk{2(p-p')}^{n-1}} \; .
\]
Multiplying all factors together
\[ \label{eqn:LL-all-loops}
\begin{split}
  \left.\bra{k\, k'} \hat{S} \ket{p\, p'}\right|_{g^n}
  & = 2^{n+1} n!
      \cdot 2^n
      \cdot \frac{1}{n!}
      \cdot \lrbrk{-\frac{ig}{2}}^n
      \cdot \frac{(-p p')^n}{\bigbrk{2(p-p')}^{n-1}}
      \cdot \frac{1}{\lambda'(p-p')} \, \delta_+(p,p',k,k') \\
  & = 2 \, \lrbrk{ig}^n \, \lrbrk{\frac{p p'}{p-p'}}^n \, \delta_+(p,p',k,k')
  \; .
\end{split}
\]
For $n=1$ we recover the tree-level result \eqref{eqn:LL-S-one-loop} computed in the main text.

\subsection{Two-particle Green's function}
\label{app:LL-instab}

Here we compute
\[
\left\langle 0\right|T\,\mathcal{O}(t,x)\mathcal{O}(0,0)
 \left|\,0\right\rangle=
 \frac{i}{4}\int_{}^{}\frac{d\Omega \, dP}{(2\pi )^2}\, e^{-i\Omega t+iPx}\,
 G(\Omega ,P) \; ,
\]
where $\mathcal{O}$ is defined in (\ref{defo}). For that we need to
compute the loop diagrams $I_0$, $I_1$ and $I_2$ off-shell:
\[
 I_0(\Omega ,P)=-\int_{}^{}\frac{d\omega \,dk}{(2\pi )^2}\,\,
 \frac{1}{(\omega -k^2)[\Omega -\omega -(P-k)^2]}
 =\int_{-\infty }^{+\infty }\frac{dk}{2\pi i}
 \,\,\frac{1}{k^2+(P-k)^2-\Omega } \; .
\]
The answer is the same (\ref{loop0})-(\ref{loop2}), where $p$, $p'$
are defined as the roots of the denominator in the integrand:
\begin{equation}\label{}
 \left\{\begin{array}{c}
 p \\ p'
 \end{array}\right\}
 =\frac{P}{2}\pm\frac{i}{2}\,\sqrt{P^2-2\Omega } \; .
\end{equation}
In particular,
\[
I_0(\Omega ,P)=\frac{1}{2i\sqrt{P^2-2\Omega }} \; .
\]
Summing the bubble diagrams as before we find:
\[
G(\Omega ,P)=\frac{4\left(pp'\right)^2}{2pp'+\frac{i}{I_0(\Omega
,P)}} \; ,
\]
which gives (\ref{2green}).

\section{Computational details for AAF model}
\label{app:AAF-details}

\subsection{Spinors}
\label{app:AAF-spinors}

In the AAF model we work with the following two-component spinors:
\[
  u(p_1) = \matr{c}{\sin \eta(p_1) \\ \cos \eta(p_1)}
  \comma
  v(p_1) = \matr{c}{\cos \eta(p_1) \\ -\sin \eta(p_1)}
\]
and their Dirac conjugates $\bar{u} = u^\dag \gamma^0$ and $\bar{v}
= v^\dag \gamma^0$. The angle $\eta(p_1)$ is defined through the
relation $\cot 2\eta(p_1) = \frac{p_1}{m}$. We use the following
explicit representation of the Dirac matrices
\[ \label{eqn:Dirac-matrices}
  \gamma^0 = \matr{cc}{0 & 1 \\ 1 & 0}
  \comma
  \gamma^1 = \matr{cc}{0 & -1 \\ 1 & 0}
  \comma
  \gamma^3 = \gamma^0 \gamma^1 = \matr{cc}{1 & 0 \\ 0 & -1} \; .
\]
These spinors obey the equations of motion
\[
  (\slashed{p}-m) u(p_1) = 0
  \comma
  (\slashed{p}+m) v(-p_1) = 0
  \; ,
\]
where $p_0 = \sqrt{p_1^2 + m^2}$. They satisfy the completeness relations
\[
  u(p_1)\bar{u}(p_1) = \frac{1}{2p_0} (\slashed{p}+m)
  \comma
  v(p_1)\bar{v}(p_1) = \frac{1}{2p_0} (\slashed{p}-m)
\]
and the orthogonality relations
\[
  u^\dag(p_1) u(p_1) = v^\dag(p_1) v(p_1) = 1
  \comma
  u^\dag(p_1) v(p_1) = v^\dag(p_1) u(p_1) = 0
\]
and
\[
  \bar{u}(p_1) u(p_1) = - \bar{v}(p_1) v(p_1) = \frac{m}{p_0}
  \comma
  u^\dag(p_1) v(p_1)  =   v^\dag(p_1) u(p_1)  = \frac{p_1}{p_0}
  \; .
\]

\subsection{Loop integrals}

We compute the ``bubble propagators'' $I_r$ in the AAF model. This is essentially a repetition of the computation in the LL model (\appref{app:LL-loop-integrals}) where now the fermion propagator \eqref{eqn:AAF-propagator} is used. Defining $\vec{P} = \vp + \vpp$ we find:
\[
\begin{split}
& \hphantom{\hspace{120mm}} \\[-3mm]
I_0(\vp,\vpp) & = \psfrag{p1}[c][c]{$p$}
                  \psfrag{p2}[c][c]{$p'$}
                  \psfrag{label1}[c][c]{$q$}
                  \psfrag{label2}[c][c]{$p+p'-q$}
                  \raisebox{-9mm}{\includegraphics*{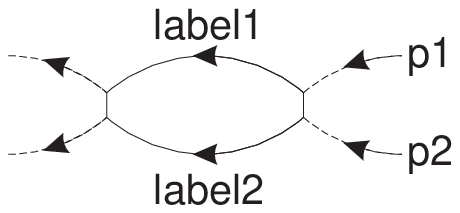}} \\[3mm]
              & = \momintt{q} D(\vq) \otimes D(\vp+\vpp-\vq) \\
              & = \frac{\vpp\times\vp}{4\vec{P}^2} \, \gamma^\alpha \otimes \gamma_\alpha
\; ,
\end{split}
\]
\[
\begin{split}
& \hphantom{\hspace{120mm}} \\[-3mm]
I_1(\vp,\vpp) & = \psfrag{p1}[c][c]{$p$}
                  \psfrag{p2}[c][c]{$p'$}
                  \psfrag{label1}[c][c]{$q$}
                  \psfrag{label2}[c][c]{$p+p'-q$}
                  \raisebox{-9mm}{\includegraphics*{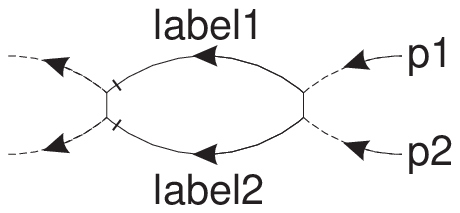}}
                  \quad = \quad
                  \psfrag{p1}[c][c]{$p$}
                  \psfrag{p2}[c][c]{$p'$}
                  \psfrag{label1}[c][c]{$q$}
                  \psfrag{label2}[c][c]{$p+p'-q$}
                  \raisebox{-9mm}{\includegraphics*{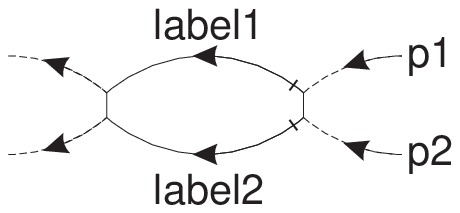}} \\[3mm]
              & = \momintt{q} \bigbrk{-\vq\times(\vp-\vq)} \, D(\vq) \otimes D(\vp+\vpp-\vq) \\
              & = \frac{\vpp\times\vp}{4\vec{P}^2} \, \cdot \Bigsbrk{
                        \sfrac{1}{2} \, \bigbrk{ \slashed{P} \gamma^3 \otimes \slashed{P} - \slashed{P} \otimes \slashed{P} \gamma^3 }
                        + m \, \bigbrk{ \slashed{P} \gamma^3 \otimes \unit - \unit \otimes \slashed{P} \gamma^3 }
                  }
\; ,
\end{split}
\]
\[
\begin{split}
& \hphantom{\hspace{120mm}} \\[-3mm]
I_2(\vp,\vpp) & = \psfrag{p1}[c][c]{$p$}
                  \psfrag{p2}[c][c]{$p'$}
                  \psfrag{label1}[c][c]{$q$}
                  \psfrag{label2}[c][c]{$p+p'-q$}
                  \raisebox{-9mm}{\includegraphics*{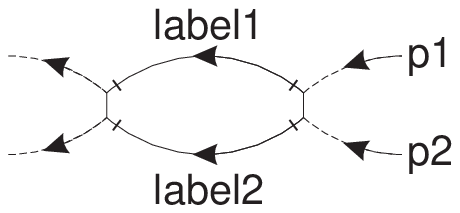}} \\[3mm]
              & = \momintt{q} \bigbrk{-\vq\times(\vp-\vq)}^2 \, D(\vq) \otimes D(\vp+\vpp-\vq) \\
              & = \frac{\vpp\times\vp}{4\vec{P}^2} \, \cdot \Bigsbrk{
                     (\vpp\times\vp)^2 \, \gamma^\alpha \otimes \gamma_\alpha
                     + m^2 \, \slashed{P} \otimes \slashed{P} \\
              & \qquad\qquad\qquad + \sfrac{m}{2} \vec{P}^2 \, \bigbrk{ \slashed{P} \otimes \unit - \unit \otimes \slashed{P} }
                     + m^2 \vec{P}^2 \, \unit \otimes \unit
                  }
\; .
\end{split}
\]
In the computation we made use of
\[
\momintt{q} \frac{1}{(q^2 + \Delta)^2} = 0 \; ,
\]
\[
\momintt{q} \frac{q^2}{(q^2 + \Delta)^2} = \frac{1}{4} \, \theta(\Delta) \; ,
\]
\[
\momintt{q} \frac{q^4}{(q^2 + \Delta)^2} = - \frac{1}{2} \Delta \, \theta(\Delta) \; ,
\]
for $\Delta\in\Reals$. These formulas can be obtained from taking derivatives of the more general integral with numerator $e^{i \vp\cdot\vec{\alpha}}$. This integral itself can be evaluated by contour integration.

\subsection{Summing all Feynman diagrams}
\label{app:AAF-higher-loops}

\begin{figure}[t]
\begin{center}
\psfrag{p1}[c][c]{$\vp$} \psfrag{p2}[c][c]{$\vpp$}
\psfrag{k1}[c][c]{$\vk$} \psfrag{k2}[c][c]{$\vkp$}
\includegraphics*{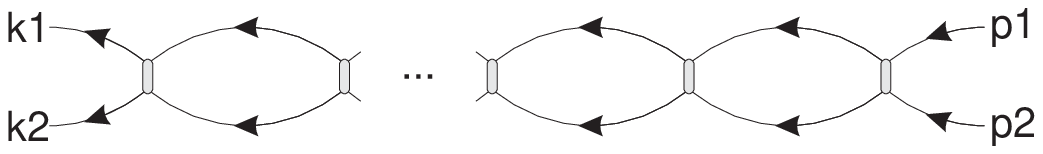}\\[3mm]
\includegraphics*{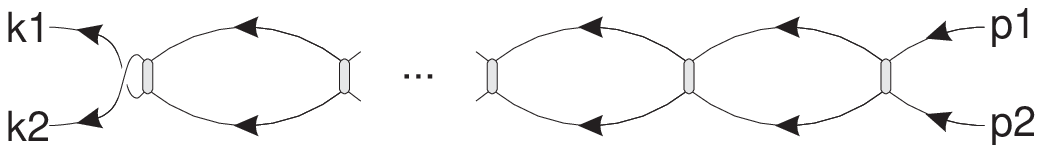}
\end{center}
\caption{\textbf{Higher loop diagrams in AAF model.}}
\label{fig:AAF-loopdiagram}
\end{figure}

We compute
\[
  \left.\bra{k\, k'} \hat{S} \ket{p\, p'}\right|_{g^n}
  = \bra{k\, k'} \, \frac{1}{n!} \lrsbrk{ \lrbrk{-\frac{ig}{m^2}} \int\!dx\,dt\,
    \varepsilon^{\alpha\beta} \bigsbrk{ \bpsi \partial_\alpha \psi \bpsi \gamma^3 \partial_\beta \psi
                                      - \partial_\alpha \bpsi \psi \partial_\beta \bpsi \gamma^3 \psi }
    }^n \ket{p\, p'} \; .
\]
For the reasons discussed in the main text only bubble diagrams
contribute. As opposed to the LL model, here the chain of bubbles is
made of two distinguished strands which connect one in-going
particle with one out-going particle each. I.e. there are two cases
(\figref{fig:AAF-loopdiagram}): $\vp$ connected to $\vk$ and $\vpp$
connected to $\vkp$ or $\vp$ connected to $\vkp$ and $\vpp$
connected to $\vk$. The strands describe the flow of the spinor
indices. At the vertex they hit onto the unit matrix or $\gamma^3$.
The two different possibilities can be taken into account by writing
the factor
\[ \label{eqn:proj}
 \Proj = \unit \otimes \gamma^3 - \gamma^3 \otimes \unit
 \comma
 \lrbrk{\half\Proj}^3 = \half\Proj
\]
at each vertex in \figref{fig:AAF-loopdiagram}.
If there was only the first vertex and if we neglect the derivatives
for a moment, then all diagrams at order $g^n$ are given by
\[
  \bigbrk{\bar{u}(k'_1)\otimes\bar{u}(k_1) - \bar{u}(k_1)\otimes\bar{u}(k'_1)} \,
  \Proj \Bigbrk{ I_0(\vp,\vpp) \Proj }^{n-1} \,
  u(p_1)\otimes u(p'_1)
  \delta_+(p_1,p'_1,k_1,k'_1) \; .
\]

Now we take into account the derivatives. The corresponding
situation in the LL model was particularly lucky as any distribution
of the derivatives led to precisely the same expression. Here, this
is not the case and we are confronted with an rather large
combinatorial problem. As all cases have to be handled separately
anyway, it turns out the be more convenient to even consider all
orders in $g$ at once. The entire higher loop series is written in
four terms:
\begin{subequations}\label{eqn:AAF-four-terms}
\begin{itemize}
\item no derivatives on external legs
\[
\begin{split}
  &
  \bigbrk{\bar{u}(k_1)\otimes\bar{u}(k'_1) - \bar{u}(k'_1)\otimes\bar{u}(k_1)} \,
  V_{10} \,
  u(p_1)\otimes u(p'_1) \,
  \delta_+(p_1,p'_1,k_1,k'_1) \\
  & =
  \bigbrk{\bar{u}(p_1)\otimes\bar{u}(p'_1) - \bar{u}(p'_1)\otimes\bar{u}(p_1)} \,
  V_{10} \,
  u(p_1)\otimes u(p'_1) \,
  \delta_-(p_1,p'_1,k_1,k'_1)
\end{split}
\]
\item derivatives only on the in-going legs
\[
\begin{split}
  &
  \bigbrk{\vpp\times\vp}
  \bigbrk{\bar{u}(k_1)\otimes\bar{u}(k'_1) - \bar{u}(k'_1)\otimes\bar{u}(k_1)} \,
  V_{11} \,
  u(p_1)\otimes u(p'_1) \,
  \delta_+(p_1,p'_1,k_1,k'_1) \\
  & =
  \bigbrk{\vpp\times\vp}
  \bigbrk{\bar{u}(p_1)\otimes\bar{u}(p'_1) - \bar{u}(p'_1)\otimes\bar{u}(p_1)} \,
  V_{11} \,
  u(p_1)\otimes u(p'_1) \,
  \delta_-(p_1,p'_1,k_1,k'_1)
\end{split}
\]
\item derivatives only on the out-going legs
\[
\begin{split}
  &
  \bigbrk{\vkp\times\vk}
  \bigbrk{\bar{u}(k_1)\otimes\bar{u}(k'_1) + \bar{u}(k'_1)\otimes\bar{u}(k_1)} \,
  V_{00} \,
  u(p_1)\otimes u(p'_1) \,
  \delta_+(p_1,p'_1,k_1,k'_1) \\
  & =
  \bigbrk{\vpp\times\vp}
  \bigbrk{\bar{u}(p_1)\otimes\bar{u}(p'_1) + \bar{u}(p'_1)\otimes\bar{u}(p_1)} \,
  V_{00} \,
  u(p_1)\otimes u(p'_1) \,
  \delta_-(p_1,p'_1,k_1,k'_1)
\end{split}
\]
\item derivatives on the in- and out-going legs
\[
\begin{split}
  &
  \bigbrk{\vpp\times\vp}
  \bigbrk{\vkp\times\vk}
  \bigbrk{\bar{u}(k_1)\otimes\bar{u}(k'_1) + \bar{u}(k'_1)\otimes\bar{u}(k_1)} \,
  V_{01} \,
  u(p_1)\otimes u(p'_1) \,
  \delta_+(p_1,p'_1,k_1,k'_1) \\
  & =
  \bigbrk{\vpp\times\vp}^2
  \bigbrk{\bar{u}(p_1)\otimes\bar{u}(p'_1) + \bar{u}(p'_1)\otimes\bar{u}(p_1)} \,
  V_{01} \,
  u(p_1)\otimes u(p'_1) \,
  \delta_-(p_1,p'_1,k_1,k'_1)
\end{split}
\]
\end{itemize}
\end{subequations}

It remains to find expressions for the $V$'s which sum all possible
diagrams. The possible placements of the derivatives along the chain
of bubbles are in one-to-one correspondence with binary sequences.
One sequence of length $n$ corresponds to one Feynman diagram at
order $g^n$. Let ``0'' stand for derivatives ``on the left'' of the
vertex (i.e. towards out-going particles) and ``1'' for derivatives
``on the right'' of the vertex (i.e. towards in-going particles).
For a diagram without derivatives on the external legs this sequence
has to begin with a ``1'' and end with a ``0'', hence the notation
$V_{10}$, etc. Now, this binary sequence translates into a sequence
of bubble propagators $I_0$ (no derivatives), $I_1$ (one pair of
derivatives) and $I_2$ (two pairs of derivatives) according to the
rules
\begin{align}
\ldots 00 \ldots & \mapsto \ldots (-I_1) \ldots \nonumber \\
\ldots 01 \ldots & \mapsto \ldots (+I_0) \ldots           \\
\ldots 10 \ldots & \mapsto \ldots (-I_2) \ldots \nonumber \\
\ldots 11 \ldots & \mapsto \ldots (+I_1) \ldots \nonumber
\end{align}
where we have chosen to include the sign from the second vertex
(Recall that the vertex with the derivatives on the left comes with
a minus sign, \eqref{eqn:AAF-vertex2}). The sign for the very first
digit in the binary sequence has to be taken into account
separately.

The next step is to generate all those sequences. The main idea is
to think of an arbitrary binary sequence as a succession of constant
subsequences, e.g. the sequence 1111000100011111000 is made of 6
constant subsequences. It is clear, that all sequences are obtained
from joining an arbitrary number of constant subsequences of
arbitrary length. The constant pieces lead to
\[ \label{eqn:constant-subsequence}
 \Proj I_\pm := \sum_{n=0}^\infty G^n (\pm \Proj I_1)^n \; ,
\]
where the minus sign is used for a sequence of zeros and the plus
sign for a sequence of ones. In \eqref{eqn:constant-subsequence} we
included for each vertex a matrix $\Proj$ and one power of the
coupling constant $G=\frac{-ig}{m^2}$. Between these constant
subsequences we have to insert a factor $G \Proj I_0$ when the
sequence changes from ``0'' to ``1'' and a factor $-G \Proj I_2$
when the sequence changes from ``1'' to ``0''. Taking all this
together we find
\begin{subequations} \label{eqn:AAF-loop-diagrams}
\begin{align}
V_{10} & := + \sum_{k=0}^\infty G^{2k+2} \Proj I_+ (-\Proj I_2) \Proj I_- \bigbrk{-\Proj I_0 \Proj I_+ \Proj I_2 \Proj I_-}^k \Proj \;,\\
V_{11} & := + \sum_{k=0}^\infty G^{2k+1} \Proj I_+                        \bigbrk{-\Proj I_2 \Proj I_- \Proj I_0 \Proj I_+}^k \Proj \;,\\
V_{00} & := - \sum_{k=0}^\infty G^{2k+1} \Proj I_-                        \bigbrk{-\Proj I_0 \Proj I_+ \Proj I_2 \Proj I_-}^k \Proj \;,\\
V_{01} & := - \sum_{k=0}^\infty G^{2k+2} \Proj I_- (+\Proj I_0) \Proj I_+ \bigbrk{-\Proj I_2 \Proj I_- \Proj I_0 \Proj I_+}^k \Proj \;.
\end{align}
\end{subequations}
The sums in \eqref{eqn:constant-subsequence} and
\eqref{eqn:AAF-loop-diagrams} can be evaluated explicitly as
functions of $\vp$, $\vpp$ and $m$. It is computationally convenient
to replace everywhere
\[
 I_k \rightarrow \tilde{I}_k = \lrbrk{\half\Proj}^2 I_k \lrbrk{\half\Proj}^2 \; .
\]
This is possible because every $I_k$ is sandwiched between two
$\Proj$'s which satisfy the property \eqref{eqn:proj}. We refrain
from printing the explicit $V$'s as the formulas are rather lengthy
and not very illuminating. However, after inserting them into
\eqref{eqn:AAF-four-terms}, evaluating the spinor products and
multiplying everything by the Jacobian $p'_0 p_0/\vpp\times\vp$ (cf.
\eqref{eqn:AAF-jacobian}) we find the fairly compact result
\[ \label{eqn:AAF-higher-loops}
  \sum_{n=1}^\infty \left.\bra{k\, k'} \hat{S} \ket{p\, p'}\right|_{g^n}
  = \frac{2 \frac{ig}{m^2} \, \vpp\times\vp}{1-\frac{ig}{m^2} \, \vpp\times\vp}
    \, \delta_-(p_1,p'_1,k_1,k'_1) \; .
\]

\section{Computational details in FR model}
\label{loopinFR}

\subsection{Loop integrals}

We need to compute
\begin{equation}\label{}
 {\rm Loop}=-\int_{}^{}\frac{d^2k}{(2\pi )^2}\,\,
 \frac{\slashed{p}+\slashed{p}'-\slashed{k}+m}{(p+p'-k)^2-m^2}
 \otimes
 \frac{\slashed{k}+m}{k^2-m^2}
\end{equation}
with the non-relativistic pole prescription.  Performing the
integral over $k^0$ first we get
\begin{equation}\label{FRloop}
 {\rm Loop}=\int_{-\infty }^{+\infty}\frac{dk}{2\pi i}\,\,
 \frac{P(k;p,p')}{2(p+p')^2(k-p_1-i\epsilon )(k-p_1'+i\epsilon )} \; .
\end{equation}
The denominator $P(k;p,p')$ is a cubic polynomial in $k$ which is
symmetric in $p$ and $p'$:
\begin{eqnarray}\label{}
 P(k;p,p')&=&2E\left[E\gamma ^0-(P-k)\gamma ^1+m\right]\otimes\left(m-k\gamma ^1\right)
 \nonumber \\ &&
 -\left[E^2+k^2-(P-k)^2\right]\gamma ^0\otimes (m-k\gamma ^1)
 \nonumber \\ &&
 +\left[E^2+k^2-(P-k)^2\right]\left[E\gamma ^0-(P-k)\gamma
 ^1+m\right]\otimes \gamma ^0
 \nonumber \\ &&
 -2E(k^2+m^2)\gamma ^0\otimes\gamma ^0 \; ,
\end{eqnarray}
where $E=p_0+p_0'$, $P=p_1+p_1'$. Its most important property is
that
\begin{eqnarray}\label{}
 P(p_1;p,p')&=&(\slashed{p}'+m)\otimes(\slashed{p}+m) \; ,
 \nonumber \\
 P(p_1';p,p')&=&(\slashed{p}+m)\otimes(\slashed{p}'+m) \; .
\end{eqnarray}
The integral in (\ref{FRloop}) diverges and requires regularization.
Any reasonable regularization (for instance dimensional
regularization) closes the contour of integration symmetrically in
the lower and upper half-planes. Evaluating the residues at $k=p_1$,
$k=p_1'$ we get eq.~(\ref{loopFR}) in the main text.



\begin{thebibliography}{99}

\bibitem{Maldacena:1998re}
J.~M.~Maldacena:
\bibtitle{The large N limit of superconformal field theories and supergravity},
Adv.~Theor.~Math.~Phys. ~\textbf{2}~(1998)~231,
\hepth{9711200}
$\bullet$
S.~S.~Gubser, I.~R.~Klebanov and A.~M.~Polyakov:
\bibtitle{Gauge theory correlators from non-critical string theory},
Phys.~Lett.~\textbf{B428}~(1998)~105,
\hepth{9802109}
$\bullet$
E.~Witten:
\bibtitle{Anti-de Sitter space and holography},
Adv.~Theor.~Math.~Phys.~\textbf{2}~(1998)~253,
\hepth{9802150}.

\bibitem{Metsaev:1998it}
R.~R.~Metsaev and A.~A.~Tseytlin:
\bibtitle{Type IIB superstring action in $\AdS_5\times\Sphere^5$ background},
Nucl.\ Phys.\ B {\bf 533}, 109 (1998),
\hepth{9805028}.

\bibitem{Bena:2003wd}
I.~Bena, J.~Polchinski and R.~Roiban:
\bibtitle{Hidden symmetries of the $\AdS_5\times\Sphere^5$ superstring},
Phys.\ Rev.\ D {\bf 69} (2004) 046002,
\hepth{0305116}.

\bibitem{Beisert:2004ry}
N.~Beisert:
\bibtitle{The dilatation operator of $\mathcal{N}=4$ super Yang-Mills theory and integrability},
Phys.\ Rept.\  {\bf 405}, 1 (2005),
\hepth{0407277}
$\bullet$
N.~Beisert:
\bibtitle{Higher-loop integrability in $\mathcal{N}=4$ gauge theory},
Comptes Rendus Physique {\bf 5}, 1039 (2004),
\hepth{0409147}
$\bullet$
J.~Plefka:
\bibtitle{Spinning strings and integrable spin chains in the AdS/CFT correspondence},
\hepth{0507136}.

\bibitem{Zarembo:2004hp}
K.~Zarembo:
\bibtitle{Semiclassical Bethe ansatz and AdS/CFT},
Comptes Rendus Physique {\bf 5}, 1081 (2004)
[Fortsch.\ Phys.\  {\bf 53}, 647 (2005)],
\hepth{0411191}.

\bibitem{Minahan:2002ve}
J.~A.~Minahan and K.~Zarembo:
\bibtitle{The Bethe-ansatz for $\mathcal{N}=4$ super Yang-Mills},
JHEP {\bf 0303}, 013 (2003),
\hepth{0212208}.

\bibitem{Beisert:2003tq}
N.~Beisert, C.~Kristjansen and M.~Staudacher:
\bibtitle{The dilatation operator of $\mathcal{N}=4$ super Yang-Mills theory},
Nucl.\ Phys.\ B {\bf 664}, 131 (2003),
\hepth{0303060}
$\bullet$
N.~Beisert and M.~Staudacher:
\bibtitle{The $\mathcal{N}=4$ SYM integrable super spin chain},
Nucl.\ Phys.\ B {\bf 670}, 439 (2003)
\hepth{0307042}
$\bullet$
N.~Beisert, V.~Dippel and M.~Staudacher:
\bibtitle{A novel long range spin chain and planar $\mathcal{N}=4$ super Yang-Mills},
JHEP {\bf 0407}, 075 (2004),
\hepth{0405001}.

\bibitem{Beisert:2005fw}
N.~Beisert and M.~Staudacher:
\bibtitle{Long-range PSU(2,2$|$4) Bethe ansaetze for gauge theory and strings},
Nucl.\ Phys.\ B {\bf 727}, 1 (2005),
\hepth{0504190}.

\bibitem{Kazakov:2004qf}
V.~A.~Kazakov, A.~Marshakov, J.~A.~Minahan and K.~Zarembo:
\bibtitle{Classical / quantum integrability in AdS/CFT},
JHEP {\bf 0405}, 024 (2004),
\hepth{0402207}.

\bibitem{Beisert:2005bm}
V.~A.~Kazakov and K.~Zarembo:
\bibtitle{Classical / quantum integrability in non-compact sector of AdS/CFT},
JHEP {\bf 0410}, 060 (2004),
\hepth{0410105}
$\bullet$
N.~Beisert, V.~A.~Kazakov and K.~Sakai:
\bibtitle{Algebraic curve for the SO(6) sector of AdS/CFT},
\hepth{0410253}
$\bullet$
S.~Sch\"afer-Nameki:
\bibtitle{The algebraic curve of 1-loop planar $\mathcal{N}=4$ SYM},
Nucl.\ Phys.\ B {\bf 714}, 3 (2005),
\hepth{0412254}
$\bullet$
N.~Beisert, V.~A.~Kazakov, K.~Sakai and K.~Zarembo:
\bibtitle{The algebraic curve of classical superstrings on $\AdS_5\times\Sphere^5$},
\hepth{0502226}.

\bibitem{Dorey:2006zj}
N.~Dorey and B.~Vicedo:
\bibtitle{On the dynamics of finite-gap solutions in classical string theory},
\hepth{0601194}.

\bibitem{Gubser:2002tv}
S.~S.~Gubser, I.~R.~Klebanov and A.~M.~Polyakov:
\bibtitle{A semi-classical limit of the gauge/string correspondence}
Nucl.~Phys.~\textbf{B636}~(2002)~99,
\hepth{0204051}
$\bullet$
S.~Frolov and A.~A.~Tseytlin:
\bibtitle{Multi-spin string solutions in $\AdS_5\times\Sphere^5$},
Nucl.~Phys.~\textbf{B668}~(2003)~77,
\hepth{0304255}
$\bullet$
S.~Frolov and A.~A.~Tseytlin:
\bibtitle{Rotating string solutions: AdS/CFT duality in non-supersymmetric sectors},
Phys.\ Lett.\ B {\bf 570}, 96 (2003),
\hepth{0306143}
$\bullet$
A.~A.~Tseytlin:
\bibtitle{Semiclassical strings and AdS/CFT},
\hepth{0409296}.

\bibitem{Bethe:1931hc}
H.~Bethe:
\bibtitle{On The Theory Of Metals. 1. Eigenvalues And Eigenfunctions For The Linear Atomic Chain},
Z.\ Phys.\ {\bf 71}, 205 (1931).

\bibitem{Faddeev:1996iy}
L.~D.~Faddeev:
\bibtitle{How Algebraic Bethe Ansatz works for integrable model},
\hepth{9605187}.

\bibitem{Korepin_book}
V.E. Korepin,  A.G. Izergin and N.M. Bogolyubov:
\bibtitle{Quantum Inverse Scattering Method, Correlation Functions and Algebraic Bethe Ansatz}
(Cambridge Univ. Press, 1992).

\bibitem{zz:factorized-s-matrix}
A.~B.~Zamolodchikov and A.~B.~Zamolodchikov:
\bibtitle{Factorized S-Matrices In Two Dimensions As The Exact Solutions Of  Certain Relativistic Quantum Field Models},
Annals Phys.\  {\bf 120} (1979) 253.

\bibitem{Rej:2005qt}
A.~Rej, D.~Serban and M.~Staudacher:
\bibtitle{Planar $\mathcal{N}=4$ gauge theory and the Hubbard model},
\hepth{0512077}.

\bibitem{Callan:2003xr}
C.~G.~Callan, H.~K.~Lee, T.~McLoughlin, J.~H.~Schwarz, I.~Swanson and X.~Wu:
\bibtitle{Quantizing string theory in $\AdS_5\times\Sphere^5$: Beyond the pp-wave},
Nucl.\ Phys.\ B {\bf 673}, 3 (2003),
\hepth{0307032}
$\bullet$
C.~G.~Callan, T.~McLoughlin and I.~J.~Swanson:
\bibtitle{Holography beyond the Penrose limit},
Nucl.\ Phys.\ B {\bf 694}, 115 (2004),
\hepth{0404007}
$\bullet$
C.~G.~Callan, T.~McLoughlin and I.~J.~Swanson:
\bibtitle{Higher impurity AdS/CFT correspondence in the near-BMN limit},
Nucl.\ Phys.\ B {\bf 700}, 271 (2004),
\hepth{0405153}
$\bullet$
T.~McLoughlin and I.~J.~Swanson:
\bibtitle{N-impurity superstring spectra near the pp-wave limit},
Nucl.\ Phys.\ B {\bf 702}, 86 (2004),
\hepth{0407240}.

\bibitem{Arutyunov:2004vx}
G.~Arutyunov, S.~Frolov and M.~Staudacher:
\bibtitle{Bethe ansatz for quantum strings},
JHEP {\bf 0410}, 016 (2004),
\hepth{0406256}.

\bibitem{Staudacher:2004tk}
M.~Staudacher:
\bibtitle{The factorized S-matrix of CFT/AdS},
JHEP {\bf 0505}, 054 (2005),
\hepth{0412188}.

\bibitem{Frolov:2006cc}
S.~Frolov, J.~Plefka and M.~Zamaklar:
\bibtitle{The $\AdS_5\times\Sphere^5$ superstring in light-cone gauge and its Bethe equations},
\hepth{0603008}.

\bibitem{Beisert:2005cw}
N.~Beisert and A.~A.~Tseytlin:
\bibtitle{On quantum corrections to spinning strings and Bethe equations},
Phys.\ Lett.\ B {\bf 629}, 102 (2005),
\hepth{0509084}.

\bibitem{Mann:2005ab}
N.~Mann and J.~Polchinski:
\bibtitle{Bethe ansatz for a quantum supercoset sigma model},
Phys.\ Rev.\ D {\bf 72}, 086002 (2005),
\hepth{0508232}.

\bibitem{Polyakov:2005ss}
A.~M.~Polyakov:
\bibtitle{Supermagnets and sigma models},
\hepth{0512310}.

\bibitem{Freyhult:2005ws}
L.~Freyhult, C.~Kristjansen and T.~M\aa{}nsson:
\bibtitle{Integrable spin chains with $U(1)^3$ symmetry and generalized Lunin-Maldacena backgrounds},
JHEP {\bf 0512}, 008 (2005),
\hepth{0510221}.

\bibitem{Beisert:2005tm}
N.~Beisert:
\bibtitle{The su(2$|$2) dynamic S-matrix},
\hepth{0511082}.

\bibitem{Faddeev:1979gh}
L.~D.~Faddeev, E.~K.~Sklyanin and L.~A.~Takhtajan:
\bibtitle{The Quantum Inverse Problem Method. 1},
Theor.\ Math.\ Phys.\  {\bf 40}, 688 (1980)
[Teor.\ Mat.\ Fiz.\  {\bf 40}, 194 (1979)].

\bibitem{Kruczenski:2003gt}
M.~Kruczenski:
\bibtitle{Spin chains and string theory},
\hepth{0311203}
$\bullet$
M.~Kruczenski, A.~V.~Ryzhov and A.~A.~Tseytlin:
\bibtitle{Large spin limit of $\AdS_5\times\Sphere^5$ string theory and low energy expansion of ferromagnetic spin chains},
Nucl.\ Phys.\ B {\bf 692}, 3 (2004),
\hepth{0403120}.

\bibitem{aaf:su11-string}
L.~F.~Alday, G.~Arutyunov and S.~Frolov:
\bibtitle{New integrable system of 2dim fermions from strings on $\AdS_5\times\Sphere^5$},
JHEP {\bf 0601}, 078 (2006),
\hepth{0508140}.

\bibitem{fr:principle-chiral-field}
L.~D.~Faddeev and N.~Y.~Reshetikhin:
\bibtitle{Integrability Of The Principal Chiral Field Model In (1+1)-Dimension},
Annals Phys.\ {\bf 167} (1986) 227.

\bibitem{Polyakov:1983tt}
A.~M.~Polyakov and P.~B.~Wiegmann:
\bibtitle{Theory Of Nonabelian Goldstone Bosons In Two Dimensions},
Phys.\ Lett.\ B {\bf 131}, 121 (1983).

\bibitem{Thacker:1980ei}
H.~B.~Thacker:
\bibtitle{Exact Integrability In Quantum Field Theory And Statistical Systems},
Rev.\ Mod.\ Phys.\  {\bf 53}, 253 (1981).

\bibitem{LLclass}
M.~Lakshmanan:
\bibtitle{Continuum spin system as an exactly solvable dynamical system},
Phys.\ Lett.\ A {\bf 61}, 53 (1977)
$\bullet$
L.A.~Takhtajan:
\bibtitle{Integration Of The Continuous Heisenberg Spin Chain Through The Inverse Scattering Method},
Phys.\ Lett.\ A {\bf 64}, 235 (1977)
$\bullet$
V.~E.~Zakharov and L.~A.~Takhtajan:
\bibtitle{Equivalence Of The Nonlinear Schr{\"o}dinger Equation And The Equation Of A Heisenberg Ferromagnet},
Theor.\ Math.\ Phys.\  {\bf 38}, 17 (1979)
[Teor.\ Mat.\ Fiz.\  {\bf 38}, 26 (1979)]
$\bullet$
R.F.~Bikbaev, A.I.~Bobenko and A.R.~Its:
\bibtitle{Finite-zone integration of the Landau-Lifshitz equation},
Dokl.\ Akad.\ Nauk SSSR {\bf 272} (1983) 1293.

\bibitem{Sklyanin}
E.K.~Sklyanin:
\bibtitle{Quantization of the Continuous Heisenberg Ferromagnet},
Lett.\ Math.\ Phys. {\bf 15} (1988) 357.

\bibitem{Thacker:1974kv}
H.~B.~Thacker:
\bibtitle{Bethe's Hypothesis And Feynman Diagrams: Exact Calculation Of A Three Body Scattering Amplitude By Perturbation Theory},
Phys.\ Rev.\ D {\bf 11}, 838 (1975)
$\bullet$
H.~B.~Thacker:
\bibtitle{Many Body Scattering Processes In A One-Dimensional Boson System},
Phys.\ Rev.\ D {\bf 14}, 3508 (1976).

\bibitem{Minahan:2005mx}
J.~A.~Minahan, A.~Tirziu and A.~A.~Tseytlin:
\bibtitle{$1/J$ corrections to semiclassical AdS/CFT states from quantum Landau-Lifshitz model},
Nucl.\ Phys.\ B {\bf 735}, 127 (2006),
\hepth{0509071}
$\bullet$
A.~Tirziu:
\bibtitle{Quantum Landau-Lifshitz model at four loops: $1/J$ and $1/J^2$ corrections to BMN energies},
\hepth{0601139}.

\bibitem{Minahan:2005qj}
J.~A.~Minahan, A.~Tirziu and A.~A.~Tseytlin:
\bibtitle{$1/J^2$ corrections to BMN energies from the quantum long range Landau-Lifshitz model},
JHEP {\bf 0511}, 031 (2005),
\hepth{0510080}.

\bibitem{Faddeev's_book}
L.D. Faddeev and L.A. Takhtajan:
\bibtitle{Hamiltonian methods in the theory of solitons}
(Springer-Verlag, 1987).

\bibitem{sutherland}
B.~Sutherland:
\bibtitle{Low-Lying Eigenstates of the One-Dimensional Heisenberg Ferromagnet for any Magnetization and Momentum},
Phys.\ Rev.\ Lett.\ {\bf 74}, 816 (1995)
$\bullet$
A.~Dhar and B.S.~Shastry:
\bibtitle{Bloch Walls and Macroscopic String States in Bethe's solution of the Heisenberg Ferromagnetic Linear Chain},
\condmat{0005397}
$\bullet$
N.~Beisert, J.~A.~Minahan, M.~Staudacher and K.~Zarembo:
\bibtitle{Stringing spins and spinning strings},
JHEP {\bf 0309} (2003) 010,
\hepth{0306139}.

\bibitem{af:su11-string}
G.~Arutyunov and S.~Frolov:
\bibtitle{Uniform light-cone gauge for strings in $\AdS_5\times\Sphere^5$: Solving su(1$|$1) sector},
JHEP {\bf 0601}, 055 (2006),
\hepth{0510208}.

\bibitem{Berezin}
F.A.~Berezin and V.N.~Sushko:
\bibtitle{Relativistic two-dimensional model of a self-interacting fermion field with nonzero mass in the state of rest},
Sov.\ Phys.\ JETP {\bf 21} (1965) 865
[Zh.\ Eksp.\ Teor.\ Fiz.\ {\bf 48} (1965) 1293].

\bibitem{Bergknoff:1978wr}
H.~Bergknoff and H.~B.~Thacker:
\bibtitle{Method For Solving The Massive Thirring Model},
Phys.\ Rev.\ Lett.\  {\bf 42}, 135 (1979)
$\bullet$
H.~Bergknoff and H.~B.~Thacker:
\bibtitle{Structure And Solution Of The Massive Thirring Model},
Phys.\ Rev.\ D {\bf 19}, 3666 (1979).

\bibitem{Korepin:1979qq}
V.~E.~Korepin:
\bibtitle{Direct Calculation Of The S Matrix In The Massive Thirring Model},
Theor.\ Math.\ Phys.\  {\bf 41}, 953 (1979)
[Teor.\ Mat.\ Fiz.\  {\bf 41}, 169 (1979)].

\bibitem{Korepin:1979hg}
V.~E.~Korepin:
\bibtitle{New Effects In The Massive Thirring Model: Repulsive Case},
Commun.\ Math.\ Phys.\  {\bf 76}, 165 (1980).

\bibitem{Mikhailov:2005sy}
A.~Mikhailov:
\bibtitle{A nonlocal Poisson bracket of the sine-Gordon model},
\hepth{0511069}.

\bibitem{Berenstein:2002jq}
D.~Berenstein, J.~M.~Maldacena and H.~Nastase:
\bibtitle{Strings in flat space and pp waves from $\mathcal{N}=4$ Super Yang Mills},
JHEP~\textbf{0204}~(2002)~013,
\hepth{0202021}.

\bibitem{4a}
N.~Gromov, V.~Kazakov, K.~Sakai and P.~Vieira:
\bibtitle{Strings as Multi-particle States of Quantum Sigma-Models},
to appear.

\end{thebibliography}
\end{document}